\def\cp{{{}_{\cal P}}}
\def\be{\begin{equation}}
\def\ee{\end{equation}}
\def\bea{\begin{eqnarray}}
\def\eea{\end{eqnarray}}
\def\nn{\nonumber}
\def\ni{\noindent}
\def\HNln{H^{(N)}_{n}}
\def\p{\partial}
\def\f{\frac}
\def\l[{\left[}
\def\r]{\right]}
\def\TG{\tilde{G}}

\def\tg{\tilde{g}}
\def\th{\tilde{h}}
\def\TXR{\tilde{X}^R}
\def\TXL{\tilde{X}^L}
\def\TXRS{\tilde{X}^{R(2)}}
\def\TXLS{\tilde{X}^{L(2)}}
\def\A{\rho^{(N)}}

\def\al{\alpha}
\def\alc{\alpha^*}

\def\vs{\varsigma}

\def\a{{\hat{a}}}
\def\ac{{\hat{a}}^{\dag}}

\def \square{\hbox{$\sqcup\!\!\!\!\sqcap$}} 
\def\w{\omega}
\def\Xsegao{\campol{a}{0}{}^{\hbox{$\!\!$2nd}}}

\def\xo{x^0{}}
\def\po{p^0{}}
\def\pomc{\frac{\po}{mc}}
\def\pmc{\frac{p}{mc}}
\def\k{\kappa}
\def\z{\zeta}

\def\overpmup{\stackrel{\leftrightarrow}{\partial}_0}

\newcommand{\parcial}[1]{ \frac{\partial}{\partial #1} }

\newcommand{\XL}[1]{ {\tilde{X}}^{L}_{#1} }

\newcommand{\XR}[1]{ {\tilde{X}}^{R}_{#1} }

\newcommand{\campol}[2]{ {\tilde{X}}^{L}_{#1 ^{#2}} }

\documentstyle[11pt,a4]{article}

\begin{document}

\begin{titlepage}

\begin{center}

{\bf QUANTUM FIELD THEORY IN A SYMMETRIC CURVED SPACE FROM 
  A SECOND QUANTIZATION ON A GROUP}\footnote{Work partially 
supported by the DGICYT.}
\end{center}

\bigskip
\bigskip

\centerline{ M. Calixto$^{1,3}$\footnote{E-mail: pymc@swansea.ac.uk / 
calixto@ugr.es},\,\, 
V. Aldaya$^{2,3}$\footnote{E-mail: valdaya@iaa.es} 
and M. Navarro$^{3}$\footnote{E-mail: mnavarro@ugr.es}}

\bigskip
\begin{enumerate}
\item {Department of Physics, University of Wales Swansea, Singleton Park, 
Swansea, SA2 8PP, U.K.}
\item {Instituto de Astrof\'{\i}sica de Andaluc\'{\i}a, Apartado Postal 3004,
18080 Granada, Spain.}
\item  {Instituto Carlos I de F\'\i sica Te\'orica y Computacional, Facultad
de Ciencias, Universidad de Granada, Campus de Fuentenueva, 
Granada 18002, Spain.} 
\end{enumerate}
\bigskip

\begin{center}

{\bf Abstract}
\end{center}

In this article we propose a ``second quantization" scheme especially 
suitable to deal with non-trivial, highly symmetric phase spaces, 
implemented within a 
more general Group Approach to Quantization, which recovers the standard
Quantum Field Theory (QFT) for ordinary relativistic linear fields. 
We emphasize, among its main virtues, greater suitability in characterizing 
vacuum states in a QFT on a highly symmetric curved 
space-time and the absence of the usual requirement
of global hyperbolicity. This can be achieved in the special case of 
the Anti-de Sitter universe, on which we explicitly construct a QFT.   

\vskip 0.5cm
\ni PACS:  04.62.+v, 03.65.Fd, 03.70.+k


\end{titlepage}


\section{Introduction}

 This paper is devoted basically to generalizing the Minkowskian 
concept of ``second quantization" to certain, non-trivial symmetric 
space-times.
By second quantization we mean the (canonical) quantization of an 
infinite-dimensional classical system constituted by the linear space of
wave functions associated with a quantum mechanical particle, the evolution 
of which is considered as a trajectory 
of a classical field. More precisely, the Fourier coefficients of the 
wave functions $a(k), a(k)^*$ are considered to be the co-ordinates of 
the phase space of the classical field to be quantized, and the 
corresponding quantum operators $\hat{a}(k), \hat{a}(k)^{\dag}$ are
interpreted as the annihilation and creation operators of identical copies
of the particle originally (firstly) quantized. The fundamental problem 
in going from Minkowski space-time to any other universe is the translation 
of the notion of annihilation and creation operators. If the mono-particle 
configuration space, where wave functions are defined, is not flat, 
there is a problem in interpreting the field operators in terms of 
annihilation and creation operators. The standard decomposition of the 
field, in flat space, into positive and negative frequency parts, $\hat{f}
=\sum_k(\phi_k \hat{a}(k)+\bar{\phi}_k\hat{a}(k)^{\dag})$, has no invariant 
meaning in curved space-time; that is, the choice of positive frequency modes 
may not be unique, hence the notion of creation and annihilation are 
inherently ambiguous \cite{Fulling,Birrell,Wald2}. This ambiguity frustrates 
any attempt to define uniquely the energy-momentum tensor by the usual method 
of normal ordering, and also implies 
a lack of a preferred definition of particles; the Hawking  effect 
\cite{Hawking2} for evaporating black holes, and the Unruh effect 
\cite{Unruh} for vacuum radiation in non-inertial reference frames,  
are a consequence of this last fact (see also \cite{conforme} for a 
connection between the Unruh effect and conformal symmetry breaking).

In general terms, the crucial point in any geometrical quantization 
procedure of a phase space with non-trivial geometry 
is the {\it global reduction} of the 
prequantization mapping (in the sense of Lie representation theory 
\cite{Kirillovb}), which is only able to account for the semiclassical or  
Bohr-Sommerfeld quantization condition. 
This reduction is achieved in practice by restricting the arguments of the
wave functions with the help of a set of partial differential equations,
usually referred to in the literature as {\it Polarization} or {\it Plank} 
conditions \cite{GQ1,GQ2,GQ3,GQ4,Kirillov}, 
leading to a given ``representation" space ($q$-space, $p$-space, 
$a^{*}$-space, etc.). The problem that then arises is to determine the 
extent to which
those restriction conditions can be consistently (globally) written.
The second quantization scheme obviously inherits the difficulty of 
globally stating the arguments of the first-quantized wave functions 
associated with the ``first-quantized" problem.

Here we propose a ``second quantization" scheme especially suitable to
deal with non-trivial, highly symmetric phase spaces, implemented 
within a more general
Group Approach to Quantization (GAQ) (see, for instance, Refs. 
\cite{GAQ,AGAQ}) , which recovers the standard Quantum
Field Theory (QFT) for known cases such as standard relativistic linear
fields. The starting point is just the group of quantum 
symmetries of the model from which GAQ extracts the corresponding dynamical 
system. 

The paper is organized as follows. In Sec. 2, we give a brief description of 
the GAQ formalism (Subsec. 2.1), as self-contained as possible, 
and particularize it for the case of Quantum Mechanics on highly 
symmetric, curved space-times. 
The simple example of the relativistic free particle, which contains most of  
all the essential elements of more general cases to which GAQ is applied, 
is worked out in Subsec. 2.2. The general approach to Quantum Mechanics 
on symmetric curved spaces of Subsec. 2.1 is illustrated with the 
specific and interesting example of the Anti-de Sitter 
universe, which is fully developed in 
Subsec. 2.3 in an unconventional picture (Fock-like picture). 
In Subsec. 3.1 we present the general scheme of ``second 
quantization"  on a group, and again use the case of the Anti-de Sitter 
universe to illustrate our procedure in Subsec. 3.2. In particular, we provide
an explicit expression for propagators and an algebraic characterization of 
vacuum states for symmetric curved spaces. Finally, Sec. 4 is devoted to a  
discussion on the restriction to the standard configuration-space picture 
and makes some comments as to how the method of GAQ differs from previous 
approaches to AdS space. In particular, we discuss how the problem of 
a lack of global hyperbolicity for the case of a symmetric curved 
space, such as the AdS space, fades away in working with the corresponding 
symmetry group.

\section{Quantum mechanics on a symmetric curved space}

In the standard approach, Quantum Mechanics on a curved space $Q$ 
is explicitly built in a configuration-space picture making use 
of the intrinsic differentiable structure of $Q$ (usually a 
globally hyperbolic pseudo-Riemanian manifold with pseudo-Riemannian 
metric $g_{\mu\nu}$). A natural generalization of this picture points towards 
the possibility of considering the space $Q$ 
as embedded within a larger differentiable structure containing  the phase 
space of the theory.  
Physically suitable for this  
is a group $G$, which will be the driver of the quantization procedure. 
We shall concentrate on the cases in which $Q$ (or, rather, its universal 
covering ${\bar{Q}}$) can be considered as (or is diffeomorphic to) 
a homogeneous space $Q=G/P$ of $G$, where $P$ is a subgroup of $G$ containing 
momentum-like coordinates $p$ and other non-dynamical (non-symplectic) 
coordinates such as rotations, gauge symmetries, etc. This case corrresponds 
to a highly symmetric curved space, although more general situations 
are being investigated in connection with some sort of ``perturbative-group''  
quantization \cite{pertur,oscipert}.  

We shall demonstrate the possibility of considering 
representation spaces (momentum, holomorphic, etc.) other than the 
standard configuration-space picture (see, for instance, the explicit 
example of the free relativistic particle in Sec. \ref{frp}). A given phase 
space may possibly be embedded in different groups. 
In this case, either the physical situation is able to select 
the appropriate group, or each of these selections might give rise to 
different quantum theories having, in particular,  non-equivalent vacua. 
This fact is partially shared with more standard approaches  to QFT 
on curved space-time \cite{Fulling,Birrell,Fulling2} (see also 
\cite{conforme}, where we discuss the different structure of the 
``Poincar\'e'' vacuum with respect to the ``conformal'' vacuum for a massless 
quantum field theory, and the effect of radiation under relativistic 
accelerations).

In working on a group, the problem of first and second canonical 
quantization on a (symmetric) curved space (which,   
relies heavily on both the structure of 
space-time itself and the structure of the 
{\it classical} solution manifold of the field equations) is translated 
to the problem of finding unitary, irreducible representations 
of a suitable group. From 
this viewpoint, the traditional and difficult problem of the lack of 
global hyperbolicity for a curved space is diluted in a group framework. 
Rather, the 
problem that really matters is of algebraic character: 
that of finding an appropriate  
polarization intended to reduce unitarily the group representation 
(see below). To clarify this situation briefly, let us remark that for a 
non-globally hyperbolic space-time, the incompleteness of the dynamics in the 
standard formulation shows up as the lack of self-adjointness of the elliptic 
operator $\hat{K}$ associated with the classical wave equation 
\cite{Fulling2,Waldjmp}
\be
(\f{\partial^2}{\partial t^2}+K)\phi=0\,.\label{K}
\ee
\ni The description of the dynamics in our group approach, however, is not 
forced to adopt this form. In fact, in a non-anomalous group 
\cite{chorri,Marmo} (non-anomalous groups correspond to those possessing an 
{\it admissible} subalgebra \cite{Kirillov}), there always exists a first-order polarization (providing a first-order differential system) which leads to 
a unitary, irreducible representation. Only when one attempts to describe, 
alternatively, the system in a configuration-space-like ``representation'' 
[paralleling (\ref{K})], does a higher-order polarization become required and 
hermiticity problems with some operators can appear. 
However, this lack of hermiticity is not directly attached to the 
lack of global hyperbolicity of space-time but, rather, to the lack of 
``classical integrability'' of the 
higher-order polarization in the sense that it does not define a proper 
classical submanifold where the wave functions have support  
(see the example of the 
Poincar\'e group in the next section and Ref. \cite{Marmo}).

Irrespective of the above mentioned  disruptive effects of 
non-globally hyperbolic space-times, there is at present increased 
interest in QFT on these spaces ---for example, QFT on 
{\it time machine universes} \cite{Hawkingtmu,Kay1}, and space-times which 
can be backgrounds for supergravity theories \cite{Peter}. 
In fact, we have chosen the case of the AdS universe to illustrate our 
method of quantization because it shares both appealing properties: 
it is highly symmetric  and non-globally hyperbolic.

\subsection{Quantization on a group $\TG$}

GAQ makes use of fibre-bundle-theory concepts, which are among the 
most powerful tools for exploring the interplay between groups and topology, 
and highlight topological quantum effects (see e.g. \cite{FracHall} for a 
relevant application).

The GAQ formalism was originally conceived \cite{GAQ} 
to improve Geometric Quantization (GQ) by freeing it from 
several limitations and technical
obstructions. Among these, we point out the impossibility of considering
quantum systems without classical limit, the lack of a proper (and naturally
defined) Schr\"odinger equation in many simple cases and the ineffectiviness 
in dealing with anomalous systems \cite{chorri,virazorro}. 

The main ingredient which enables GAQ to avoid these limitations is a Lie group
structure on the manifold $\tilde{G}$ replacing the quantum manifold $Q_P$ of
GQ. $\tilde{G}$ is also a principal bundle with structure group $U(1)$, but now
$\tilde{G}/U(1)$ is not forced to wear a symplectic structure. In this way, 
non-symplectic parameters associated with 
symmetries such as time translations, 
rotations, gauge transformations, etc. are naturally allowed and give rise to 
relevant operators (Hamiltonian, angular momentum, null charges,
etc.). In addition, on any Lie group, there are always two sets of mutually 
commuting vector fields. In fact, the sets of left- and right-invariant 
vector fields constitute two realizations of the Lie algebra of the group, one 
of which can be used to represent the group, and the other to reduce 
the representation in a compatible manner (see below).

Needless to say, the requirement of a group structure in $\tilde{G}$ 
represent some drawback, although it is lesser, in practice, 
than it might seem. After all, any consistent (non-perturbative) 
quantization is only a unitary irreducible representation of a 
suitable (Lie, Poisson) algebra. Also, constrained quantization (see below
and Refs. \cite{AGAQ,FracHall}) increases the range of applicability of the 
formalism.

Nonetheless, we  should remark that the GAQ 
formalism is not meant to quantize
a classical system (a phase space) but, rather, 
the quantizing group is the primary quantity and in some 
cases (anomalous groups \cite{chorri,Marmo,virazorro}, for instance) 
it is unclear how to
associate a phase space with the quantum theory obtained.

The starting point of GAQ is a group $\TG$ (the quantizing group) 
with a principal fibre bundle 
structure $\TG(M,T)$, having $T$ as the structure group and $M$ being 
the base. The group $T$ generalizes 
the phase invariance of Quantum Mechanics. Although the situation can be more 
general \cite{AGAQ}, we shall
start with the rather general case in which $\TG$ is a central extension of
a group $G$ by $T$ [$T=U(1)$ or even $T=C^*=\Re^+\times U(1)$]. 
For the one-parametric group 
$T=U(1)$, the group law for $\TG=\{\tilde{g}=(g,\zeta)/ g\in G, 
\zeta\in U(1)\}$ adopts the following 
form:
\be
\tilde{g}'*\tilde{g}=(g'*g,\zeta'\zeta e^{i\xi(g',g)})\;,\label{gtlaw}
\ee
\ni where $g''=g'*g$ is the group operation in $G$ and $\xi(g',g)$ is a 
two-cocycle of $G$ on $\Re$ fulfilling:
\be
\xi(g_2,g_1)+\xi(g_2*g_1,g_3)=\xi(g_2,g_1*g_3)+\xi(g_1,g_3)\;\;, g_i\in G. 
\ee
\ni In the general theory of central extensions \cite{Extensiones},  
 two-cocycles are said to be equivalent if they differ in a two-coboundary, 
i.e. a two-cocycle which can be written in the form  $\xi(g',g)=\delta(g'*g)-
\delta(g')-\delta(g)$, where $\delta(g)$ is 
called the generating function of the two-coboundary (from now on we shall 
omit the prefix ``two'' when referring to both cocycles and coboundaries). 
However, although cocycles 
differing on a coboundary lead to equivalent central extensions as such, 
there are some coboundaries which provide a non-trivial connection on the 
fibre bundle $\TG$ and Lie-algebra structure constants different from that 
of the direct product $G\times U(1)$. These are generated by a function 
$\delta$ with a non-trivial gradient at the 
identity, and can be divided into Pseudo-cohomology equivalence subclasses: 
two pseudo-cocycles are equivalent if they differ in a coboundary generated 
by a function with trivial gradient at the identity 
\cite{Saletan,Pseudoco,Marmo}. Pseudo-cohomology plays 
an important role in the theory of 
finite-dimensional semi-simple groups, as they have trivial cohomology. For 
them, Pseudo-cohomology classes are associated with coadjoint orbits 
\cite{Marmo2}.

The right and left finite actions of the group $\TG$ on 
itself provide two sets 
of mutually commuting (left- and right-, respectively) invariant vector fields:
\be
\TXL_{\tg^i}=\left.\f{\p {\tg''}{}^j}{\p \tg^i}\right|_{\tg=e}
\f{\p}{\p \tg^j},\;\;\;\TXR_{\tg^i}=\left.
\f{\p {\tg''}{}^j}{\p {\tg'}{}^i}\right|_{\tg'=e}\f{\p}{\p \tg^j},\;\;\; 
\l[\TXL_{\tg^i},\TXR_{\tg^j}\r]=0,\label{txlr1}
\ee
\ni where $\{\tg^j\}$ is a parameterization of $\TG$.
The GAQ program continues by finding the left-invariant 1-form $\Theta$ (the 
{\it Quantization 1-form})
associated with the central generator $\TXL_\zeta=\TXR_\zeta, \zeta\in T$; 
that is, the $T$-component
$\tilde{\theta}^{L(\zeta)}$ of the canonical left-invariant 1-form
$\tilde{\theta}^L$ on $\tilde{G}$. This constitutes the generalization of the 
Poincar\'e-Cartan form of Classical Mechanics (see \cite{Abraham}).
The differential $d\Theta$ is a
{\it presymplectic} form and its {\it characteristic module}, $\hbox{Ker}
\Theta\cap\hbox{Ker} {}d\Theta$, is generated by a left subalgebra 
${\cal G}_\Theta$ named the {\it
characteristic subalgebra}. The quotient $(\TG, \Theta)/{\cal G}_\Theta$ is a
{\it quantum manifold} in the sense of Geometric Quantization 
\cite{GQ1,GQ2,GQ3,GQ4}. The trajectories generated by the vector fields  in 
${\cal G}_\Theta$ constitute the generalized equations of motion of 
the theory (temporal evolution, rotations, gauge symmetries, etc.), 
and the Noether 
invariants under those equations are $F_{\tg^j}\equiv i_{\TXR_{\tg^j}}\Theta$;
that is, the contraction of right-invariant vector fields with 
the Quantization 1-form. Vector fields with null Noether invariant are 
called {\it gauge} and close an {\it horizontal ideal} of the whole Lie 
algebra of $\TG$ (see Ref. \cite{config2}).

Let ${\cal B}(\TG)$ be the set of complex-valued
$T$-{\it functions} on $\TG$ in the sense
of principal bundle theory: 
\be
\psi(\zeta*\tg)=D_{T}(\zeta)\psi(\tg), \;\;\zeta\in T\label{tcondition}\;,
\ee
\ni where $D_{T}$ is the natural representation of $T$ 
on the complex numbers $C$. 
The representation of $\TG$ on ${\cal B}(\TG)$ generated
by ${\tilde{\cal G}}^R=\{\TXR\}$ is called {\it Bohr Quantization} and is 
{\it reducible}. The reduction can be achieved by means of left restrictions 
on wave functions $\psi$ compatible with (\ref{tcondition}); that is, 
by imposing a {\it full polarization} ${\cal P}$:
\be
\TXL\psi_\cp=0,\;\;\forall \TXL\in {\cal P}\label{defpol}\;,
\ee
\ni which is a maximal, horizontal (excluding $\TXL_\zeta$) left
subalgebra of ${\tilde{\cal G}}^L$ which contains ${\cal G}_\Theta$. There 
is a one-to-one correspondence between full polarizations and classes of 
inequivalent  irreducible representations (physical systems). This problem was 
firstly studied by A.A. Kirillov (Ref. \cite{Kirillov}) under the different 
denomination of ``admissible subalgebras''.

It should be noted that the existence of a full polarization, containing 
the whole subalgebra ${\cal G}_\Theta$, is not guaranteed. In case of such a
breakdown, called an {\it anomaly}, or simply  the desire to choose 
a preferred representation space, a higher-order polarization  
${\cal P}^{HO}$ 
must be imposed \cite{chorri,Marmo}. A higher-order polarization 
is a maximal, horizontal subalgebra of the left enveloping algebra 
$U\tilde{{\cal G}}^L$ which contains ${\cal G}_\Theta$.
The kind of 
theory that interests us is a particular case of this last situation; 
for the case of a representation space $Q=G/P\sim(t,\vec{x})$, 
the higher-order polarization 
must be made of the extended vector fields corresponding to the Lie 
algebra of the subgroup $P$ (which is parametrized by the co-ordinates $h^i$ 
complementary to the space-time ones, i.e., boosts, rotations, 
gauge symmetries, etc) 
and a ``deformation" ${\tilde X}^{HO}_t$ of the vector field 
$\TXL_t$ associated with the temporal evolution which, for most cases, 
can be chosen to be a Casimir operator of $G$. In summary,
\be
{\cal P}^{HO}=<{\tilde X}^{HO}_t,\TXL_{h^i}>,\;\;  h^i\in P\;.\label{defpol2}
\ee
\ni If the group contains more than one Casimir operator, the extension 
procedure $G\rightarrow \TG$ chooses one of them and the higher-order 
polarization condition ${\tilde X}^{HO}_t\psi=0$ will represent the 
equation of motion of the theory (a generalization of the 
relativistic wave equations: Klein-Gordon, Dirac, etc.,  when the 
curved space $Q$ is not locally Lorentzian).

The group $\TG$ is irreducibly represented on the space 
${\cal H}(\TG)\equiv\left\{|\psi\rangle \right\}$ of polarized wave 
functions, and on its dual  
${\cal H}^*(\TG)\equiv\left\{\langle \psi|\right\}$. If we denote by 
\be
\psi_\cp(\tg)\equiv\langle \tg_\cp|\psi\rangle \,,\,\,
\psi'{}^*_\cp(\tg)\equiv\langle \psi'|\tg_\cp\rangle 
\ee
\ni the coordinates of the ``ket" $|\psi\rangle $ and the ``bra" 
$\langle \psi'|$ in a representation defined through 
a polarization ${\cal P}$ (first- or higher-order), then, a scalar product 
on ${\cal H}(\TG)$ can be naturally defined as:
\be
\langle \psi'|\psi\rangle \equiv
\int_{\TG}{ \mu(\tg)\psi'{}^*_\cp(\tg)\psi_\cp(\tg)},
\ee
\ni where  
\be
 \mu(\tg)\equiv\theta^L_{\tg^1}\wedge
\stackrel{\hbox{dim}(\TG)}{\dots}\wedge\theta^L_{\tg^n}\label{liiv}
\ee
\ni is the left-invariant integration volume in $\TG$ and
\be
1=\int_{\TG}{|\tg_\cp\rangle  \mu(\tg)\langle \tg_\cp|}\label{closure}
\ee
\ni formally represents a {\it closure} relation.
A direct computation proves that, with this scalar product, 
the group $\TG$ is unitarily represented through the 
{\it left} finite action ($\rho$ denotes the representation)
\be
\langle \tg_\cp|\rho(\tg')|\psi\rangle \equiv\psi_\cp(\tg'{}^{-1}*\tg) 
\label{leftaction}
\ee
\ni The {\it adjoint} action is then defined as 
\be
\langle \psi'|\rho^{\dag} (\tg')|\psi\rangle\equiv
\langle \psi|\rho(\tg')|\psi'\rangle^* ,\;\;\hbox{i.e}\;\;
\langle \tg_\cp|\rho^{\dag} (\tg')|\psi\rangle =\psi_\cp(\tg'*\tg)
\,.\label{adjoint}
\ee

We can relate the coordinates of $|\psi\rangle $ in two  given
representations, corresponding to two different polarizations  ${\cal
P}_1$ and ${\cal P}_2$, as follows 
\be 
\psi_{\cp_1}(\tg)=\langle
\tg_{\cp_1}|\psi\rangle = \int_{\TG}{ \mu(\tg')\langle
\tg_{\cp_1}|\tg_{\cp_2}'\rangle  \langle \tg_{\cp_2}'|\psi\rangle
}\equiv\int_{\TG}{ \mu(\tg')
\Delta_{\cp_1\cp_2}(\tg,\tg')\psi_{\cp_2}(\tg')}\,,\label{polchange}
\ee 

\ni where $\Delta_{\cp_1\cp_2}(\tg,\tg')$ is a
``polarization-changing operator'' (for example, 
the Fourier transform $\langle x| p\rangle$, the Bargmann transform 
$\langle x| a\rangle$, etc). The finiteness of the 
integration (\ref{polchange}) for non-compact semi-simple groups is 
controlled by the partial weights in the wave functions, as a 
consequence of the polarization equations, which damps the wave functions 
down  for specific values of the pseudo-cohomology extension parameters 
---see later on Eq. (\ref{wavepol}) for the specific case of $SU(1,1)$ 
and Ref. \cite{Hermann}. For non-semisimple and non-compact 
(but non-anomalous) groups, $i_{\XL{h^1}} i_{\XL{h^2}}\dots i_{\XL{h^j}}
\mu(\tg),\,\,\XL{h^i}\in {\cal P}^{HO}\,,$ is an invariant 
restricted measure on $Q$. 
However, a minimal representation on a Cauchy hypersurface $\Sigma\subset Q$ 
would require a regularization  procedure for each particular case.

An explicit expression of $\Delta_{\cp_1\cp_2}$ is possible by using 
a basis ${\cal B}({\cal H}(\TG))=\left\{|n\rangle \right\}_{n\in I}$ 
($I$ is a set of index, non-necessarily discreet) of 
${\cal H}(\TG)$, as follows
\be
\Delta_{\cp_1\cp_2}(\tg,\tg')=\langle \tg_{\cp_1}|\tg_{\cp_2}'\rangle
= \sum_{n\in I}\psi_{\cp_1,n}^*(\tg)\psi_{\cp_2,n}(\tg')\,,\label{operchan} 
\ee 
\ni where $\psi_{\cp_i,n}(\tg)\equiv\langle \tg_{\cp_i}|n\rangle $ are the 
coordinates of $|n\rangle $ in a polarization ${\cal P}_i$. The basis 
$\left\{|n\rangle \right\}_{n\in I}$ is made of eigenvectors of the right 
counterpart of an Abelian subalgebra of ${\cal P}$; in particular, of 
the Hamiltonian $\TXR_t$. 

{\it Constraints} are consistently incorporated into the theory by 
enlarging  the structure group $T$ (which always includes $U(1)$), i.e, 
through  $T$-function conditions:
\be
\rho(\tilde{\tau})|\psi\rangle=
D^{(\epsilon)}_T(\tilde{\tau})|\psi\rangle\,,\;\;\tilde{\tau}\in T 
\ee
\ni or, for continuous transformations,
\be
\TXR_{\tilde{\tau}}|\psi\rangle =
dD^{(\epsilon)}_T(\tilde{\tau})|\psi\rangle \;, \label{const}
\ee 
$D^{(\epsilon)}_T$ means a specific representation of $T$  [the index 
$\epsilon$ parametrizes different (inequivalent) quantizations] and 
$dD^{(\epsilon)}_T$ is its differential. As a particular example, 
let us mention the case when Quantum Mechanics on $Q$ (non-simply connected) 
is recovered 
from Quantum Mechanics on its universal covering $\bar{Q}$ by 
choosing $T=\Pi_1(Q)\times U(1)$ as the structure group [$\Pi_1(Q)$ means 
the first homotopy group of $Q$], thus 
leading to topological quantum effects commonly known as 
$\vartheta$-structure \cite{Jackiw} (see also Ref. \cite{FracHall}).  

It is clear that, for a non-central structure 
group $T$, not all the right operators $\TXR_{\tg}$ will preserve 
these constraints; a sufficient condition for a 
subgroup  $\TG_T\subset\TG$ to preserve the constraints is 
(see \cite{FracHall}):
\be
 \l[\TG_T,T\r]\subset \hbox{Ker}D^{(\epsilon)}_T
\ee
\ni  [note that, for the trivial representation of $T$, the subgroup 
$\TG_T$ is simply the {\it normalizer} of $T$]. 
$\TG_T$ belongs to the set of {\it good} operators \cite{AGAQ},  for which 
the subgroup $T$ behaves as a {\it gauge} group (see \cite{config2} for a 
thorough study of gauge symmetries and constraints from the standpoint of 
GAQ). A more general situation can be posed in which the 
constraints are lifted to higher-order level, not necessarily 
first-order as in (\ref{const}); that is, when the constraints 
constitute a subalgebra of the 
right enveloping algebra $U\tilde{{\cal G}}^R$. A useful example of this 
last case results when we select representations 
labelled by a value $\epsilon$ 
of some Casimir operator $K$ of a subgroup $\TG_K$ of $\TG$ 
(see later on in Sec. 3.1 and Ref. \cite{conforme} 
for a relevant application). The good operators have to be found, in general, 
in the right enveloping algebra with the only condition of preserving 
the constraint. 

In a more general case, in which $T$ is not a trivial central extension,
$T \neq \check{T} \times U(1)$, where $\check{T} \equiv T/U(1)$ ---i.e. $T$ 
contains second-class constraints--- the conditions (\ref{const}) are not all
compatible and we must select a subgroup $T_B = T_p \times U(1)$, where
$T_p$ is the subgroup associated with a right polarization subalgebra of the
central extension $T$ (see \cite{AGAQ}).
   
For simplicity, we have sometimes made use of infinitesimal (geometrical) 
concepts, but all of this language can be translated 
to their finite (algebraic) 
counterparts (see \cite{AGAQ}), a desirable way of proceeding when discrete  
transformations are incorporated into the theory.

\subsection{Quantization of the free relativistic particle\label{frp}}

Let us illustrate the abstract construction above with the help of the 
simple example of the free relativistic particle. 
Our starting point
is the group law for the ordinary (non-extended) Poincar\'e group $G$ in 1+1D 
(see \cite{Position} and
references therein for the 1+3D case).  
It is easily derived from its action on the 1+1D Minkowski space-time
parametrized by $\{a^\mu\} \equiv\{a^0,a^1=a\}$:
 $a'^\mu=\Lambda^\mu_{.\nu}(p^0,p)a^\nu + x^\mu$, where $\{x^\mu\}$ 
are the translations
and $\Lambda$, the boosts, are parametrized by
either $p$ or 
\be 
\chi\equiv \sinh^{-1}\frac{p}{mc}\equiv \sinh^{-1}(\gamma 
\frac{V}{c})\,.\label{lova}
\ee
\ni  $\chi$ is the hyperpolar co-ordinate 
parametrizing the (upper sheet of the) 
hyperboloid $\po{}^2-p^2=m^2c^2$, 
often referred to as the  {\it Lobachevsky space}\  
(see \cite{Mir-Kasimov3} and references therein). In terms of $p$, we have 
\be 
\Lambda =\left(\begin{array}{cc} \frac{p^0}{mc}&\frac{p}{mc} \\
 \frac{p}{mc}& 1+\frac{p^2}{mc(p^0+mc)}\end{array}\right)\,.
\ee
\ni As a manifold,
the group can be seen as the direct product of Minkowski space-time and 
the mass hyperboloid. 

The consecutive action of two Poincar\'e transformations leads to the 
composition law
\begin{eqnarray}
\xo'' &=& \xo' + \frac{\po'}{mc}\, \xo + \frac{p'}{mc}\, x \,,\nn \\
x'' &=& x' + \frac{\po'}{mc}\, x + \frac{p'}{mc}\, \xo\,, \label{Poincare} \\
p'' &=& \frac{\po}{mc}\, p' + \frac{\po'}{mc}\, p\,. \nn 
\end{eqnarray}  

The Poincar\'e group admits only trivial central extensions by $U(1)$,
i.e. extensions of the form (\ref{gtlaw}) 
where the cocycle $\xi$ is a { coboundary} generated by a 
function $\delta$ on $G$, $\xi(g',g)=\delta(g'*g)-\delta(g')-\delta(g)$.
We choose $\delta(g)=mcx^0$, so that the $U(1)$ law to be added to 
(\ref{Poincare}) is
\begin{equation}
\zeta''=\zeta'\zeta e^{imc(\xo''-\xo'-\xo)}\,. \label{U(1)}
\end{equation}

>From (\ref{Poincare}) and (\ref{U(1)}) we immediately derive both
left- and right-invariant vector fields:

\begin{eqnarray}
\XL{\xo} &=& \frac{\po}{mc}\parcial{\xo} + \frac{p}{mc}\parcial{x} +
            \frac{\po(\po-mc)}{mc}i\z\parcial{\z}\,, \nn \\
\XL{x} &=& \pomc\parcial{x} + \pmc\parcial{\xo} + \frac{p(\po-mc)}{mc}
i\z\parcial{\z}\,,  \nn \\
\XL{p} &=& \pomc\parcial{p} + \pomc\, xi\z\parcial{\z}\,, \nn \\
\XL{\z} &=& \XR{\z}=\z\parcial{\z}\,, \\
\XR{\xo} &=& \parcial{\xo}\,, \nn \\
\XR{x} &=& \parcial{x} + pi\z\parcial{\z}\,, \nn \\
\XR{p} &=& \pomc\parcial{p} + \frac{\xo}{mc}\parcial{x} + 
        \frac{x}{mc}\parcial{\xo} + \left( \pmc\,\xo + \pomc\,x-x\right)
i\z\parcial{\z}\,. \nn 
\end{eqnarray}

\ni The pseudo-extended Poincar\'e algebra become

\begin{eqnarray}
\left[\XR{\xo}, \XR{x} \right] &=& 0 \,,\nn \\
\left[\XR{\xo}, \XR{p} \right] &=& \frac{1}{mc}\XR{x}\,,  \\
\left[\XR{x}, \XR{p} \right] &=& \frac{1}{mc}\XR{\xo} - i\XR{\z}\,. \nn
\end{eqnarray}

\ni Notice the appearance of the central generator $\XR{\z}$ in the third 
commutator above, making the extension by $U(1)$ less trivial  and
justifying the name of pseudo-extension. 

The Quantization 1--form and the Characteristic Module are:
\bea
\Theta&=& -(p^0-mc)dx^0-xdp-i\z^{-1} d\z\,,\\
{\cal G}_\Theta&=&<\XL{\xo}>\,.\nn
\eea
\ni The pseudo-extended Poincar\'e group admits a first-order full 
polarization ${\cal P}_p$, which is generated by 
\be
{\cal P}_p=<\XL{\xo},\XL{x}>\,,
\ee
\ni  leading to the momentum representation.  
The corresponding
polarized
$U(1)$-functions (\ref{tcondition}), $\XR{\zeta}\psi=\psi$,  are 
$\psi_{\cp_p} = \zeta\exp[-i(\po-mc)\xo]\phi(p)$, 
and the right generators act on them
as quantum operators:

\begin{equation}
\begin{array}{rlll}
\hat{p}^0\,\psi_{\cp_p} \equiv&i(\XR{\xo}+imc\XR{\z})\psi_{\cp_p}&
=\po\psi_{\cp_p} &\Rightarrow\hat{p}^0\,\phi(p)=\po\,\phi(p) \,,\\
\hat{p}\,\psi_{\cp_p} \equiv&-i\XR{x}\psi_{\cp_p}&=p\psi_{\cp_p} &
\Rightarrow\hat{p}\,\phi(p)=p\phi(p)\,, \\
\hat{k}\,\psi_{\cp_p} \equiv&i\XR{p}\psi_{\cp_p}&= \zeta i\pomc\,
e^{-i(\po-mc)\xo}\frac{\partial\phi}{\partial p} &\Rightarrow\hat{k}\,\phi(p)= 
i\pomc \frac{\partial\phi(p)}{\partial p}\,.\end{array}\label{reprep}
\end{equation}
In this identification of quantum operator with right generators the
rest mass energy has been added to the time generator to obtain the
true energy operator $\hat{p}^0$. This is a consequence  of the fact that 
the pseudo-extension is simply a redefinition of the $U(1)$
parameter. The representation (\ref{reprep}) is unitary with the 
natural measure (\ref{liiv}) restricted to the momentum space: 
\begin{equation}
\mu(\tg) = -i\frac{mc}{\po} d\xo\wedge dx\wedge dp\wedge \zeta^{-1}d\zeta
\rightarrow i_{\XL{\zeta}}i_{\XL{x}}i_{\XL{x^0}}\mu(\tg)=
\frac{mc}{\po} dp\,.\label{pliiv}
\end{equation}
\ni We should realize, however,  that the boost operator $\hat{k}$ is 
not a true position
operator, i.e. it is not $i\frac{\partial}{\partial p}$ and does not generate
ordinary translations in the spectrum of the momentum operator $\hat{p}$.

Unfortunately, there is no first-order full polarization leading 
to the configuration-space representation. In fact, $\XL{p}$ and $\XL{\xo}$ 
 do not close a proper horizontal subalgebra. However, 
we can resort to a higher-order polarization (\ref{defpol2}) 
including the $\XL{p}$ generator: 
\be
{\cal P}_x^{HO} = < \XL{\xo}{}^{HO}\equiv 
\XL{\xo} + i\left[ \sqrt{m^2c^2-(\XL{x})^2} -
mc\right]\XL{\z},\,\, \XL{p} > \,. \label{higo}
\ee
\ni We should note in passing that the infinite-order character of 
$\XL{\xo}{}^{HO}$ is due to the restriction to the upper 
sheet of the mass hyperboloid
and that a second-order operator exists, 

\begin{equation}
\Xsegao = \campol{x}{0} + \frac{i\hbar}{2mc}\left[\left(\campol{x}{0}\right)^2
                                    -\left(\campol{x}{}\right)^2\right],
\label{KGEQN}
\end{equation}

\noindent which leads to the Klein-Gordon equation. We can say that the
polarization containing  $\XL{\xo}{}^{HO}$ gives a highest-weight 
representation, whereas that containing $\Xsegao$ gives a 
(reducible, since it contains the two signs of the energy) 
representation characterized by a given value (the mass) of the Casimir 
operator of the group. This duality is also valid for more general 
non-compact Lie groups. 

By solving the polarization equations associated with (\ref{higo}), we arrive 
to the general expression for wave functions
\be
\psi_{\cp_x}=\zeta\exp(-ixp)\exp\left\{-imc\xo 
\left(\sqrt{m^2c^2-\frac{\partial^2}{\partial x^2}}-mc\right)\right\}\,
\phi(x)\,.\label{wahigo}
\ee
\ni Since the form of the polarized wave functions (\ref{wahigo}) is 
preserved by the right-invariant vector fields, it makes sense to restrict 
the operators to the arbitrary functions $\phi(x)$. The resulting operators 
are:
\begin{eqnarray}
\hat{p}^0\phi(x) &=&\sqrt{m^2c^2-\frac{\partial^2}{\partial x^2}}\,\phi(x)\nn\\
\hat{p}\phi(x) &=& -i\parcial{x}\,\phi(x) \label{unitpoin} \\
\hat{k}\phi(x) &=& \left[\frac{x}{mc}\sqrt{m^2c^2-
\frac{\partial^2}{\partial x^2}} \right]\,\phi(x) \,.\nn
\end{eqnarray}

\ni Unlike the non-relativistic configuration-space representation, this 
{\it minimal} realization (restricted to $x$ and $\parcial{x}$) 
is not unitary with the trivially regularized restriction of the 
measure (\ref{pliiv}), $\mu(\tg)\rightarrow dx$, 
even though the representation is unitary on the complete 
wave functions, because the $\hat{k}$ operator is not hermitian. This 
breakdown is a direct consequence of the {\it weak} closure of the 
higher-order polarization (\ref{higo}). Although it closes on polarized 
wave functions, thus giving a well-defined, irreducible carrier subspace, the 
polarization itself is not integrable in the classical sense. The transverse 
space to the polarization subalgebra, that which should be the $x$-space, 
is not properly defined in the weak case. Nonetheless, the transformation 
$\phi(x)\equiv e^{-ix^0\hat{p}^0}\sqrt{\hat{p}^0}\varphi(x,x^0)$ restores 
unitarity, taking the boost operator to the symmetrized form 
\be
\frac{1}{\sqrt{\hat{p}^0}}\,\hat{k}\sqrt{\hat{p}^0}=\frac{1}{2}
\left(\frac{x}{mc}\sqrt{m^2c^2-\frac{\partial^2}{\partial x^2}}+
\sqrt{m^2c^2-\frac{\partial^2}{\partial x^2}}\frac{x}{mc}\right)\,.
\ee 
\ni In fact, this transformation agrees with the standard prescription 
for the scalar product of relativistic fields: $\int{dx \phi^*(x)\phi'(x)}
=\frac{1}{2}\int{dx \varphi^*(x,x^0)\overpmup \varphi'(x,x^0)}$. 

Another way of looking at the unitarity problem of minimal configuration-space 
representations consists of finding a {\it strongly} closing higher-order 
polarization. This can be achieved, in the case of the Poincar\'e group, 
by adding a new momentum operator $\hat{\pi}$ and a new central extension 
reproducing a {\it canonical} commutation relation 
$[\hat{k},\hat{\pi}]=i\hat{1}$, thus leading  to  a  
finite-dimensional enlarged 
Poincar\'e algebra: the S-Poincar\'e algebra \cite{spoincare}. 
The operator $\hat{\pi}\equiv i\XR{\k}$  
generates true translations on the spectrum of $\hat{k}$, $\k$, the extra  
parameter of the group. The spectrum of
$\hat{\pi}$, $\pi\equiv mc\chi$, is related to $p$ through (\ref{lova}). The 
explicit expression for the abovementioned higher-order polarization is:
\be
{\cal P}_\k^{HO} = < \XL{\xo}+imc\left[\cosh (\frac{i}{mc}\XL{\k}) -1\right]
\XL{\z},\,  \XL{x}+imc\,\sinh (\frac{i}{mc}\XL{\k})\,\XL{\z},\,\XL{p} >\,.
\ee
\ni The resulting minimal configuration-space representation proves to be:
\begin{eqnarray}
\hat{p}^0\phi(\k) &=& mc \cosh(\frac{-i}{mc}\parcial{\k})\,\phi(\k)\,, \nn\\
\hat{p}\phi(\k) &=& mc \sinh(\frac{-i}{mc}\parcial{\k})\,\phi(\k)\,, \nn \\
\hat{k}\phi(\k) &=& \k\,\phi(\k)\,,  \\
\hat{\pi}\phi(\k) &=& -i \parcial{\k}\,\phi(\k)\,, \nn
\end{eqnarray}
\ni which is unitary with respect to the restricted measure 
$d\k$ (see \cite{spoincare} 
for more details).

\subsection{Quantum mechanics on the Anti-de Sitter space-time\label{holosec}}

Anti-de Sitter space-time (AdS) \cite{Hawking,Weinberg} 
in 1+1 dimensions can be seen as a homogeneous 
space of the group $G=SO(1,2)$ ($SO(3,2)$ in 3+1 dimensions). It is then a 
particular case of curved space on which a group quantization method is 
especially  suited. We must remark that, in restricting to 1+1 dimensions, we 
find an apparent ambiguity as $SO(1,2)\sim SO(2,1)$. However, the distinction
between de Sitter and Anti-de Sitter spaces in 1+1 dimensions is realized 
by means of two different central extensions, associated with non-equivalent
co-adjoint orbits (see \cite{conforme} for two different pseudo-extensions of 
$SO(1,2)$; see also \cite{Marmo} for a general discussion on 
pseudo-extensions and co-adjoint orbits). Furthermore, and except 
for discrete transformations, which
are not relevant for our purpose, $G$ can be replaced by its two-covering  
\be
SU(1,1)=\left\{ U= \left( \begin{array}{cc} z_1&z_2\\z_2^*&z_1^*\end{array}
\right) ,z_i,z_i^* \in C/ \det(U)=|z_1|^2-|z_2|^2=1 \right\}\;,
\ee
\ni  which is more directly related to the representation 
(and central extension) we are going to handle. The group $SU(1,1)$ 
has the topological structure of a trivial fibre bundle 
with fibre $U(1)$ and base the hyperboloid (or its projection on the 
plane: the open unit disk $D_1$). A system of coordinates 
adapted to this fibration is the following:

\be
\eta\equiv\f{z_1}{|z_1|}, \;\; \al\equiv\f{z_2}{z_1},\;\;
\al^*\equiv\f{z_2^*}{z_1^*},\;\;\;\; \eta\in U(1),\;\;\al,
\al^*\in D_1\;,\label{ycero}
\ee
\ni where $y_0\equiv -i\log\eta$ will play the role of a (dimensionless) 
time coordinate and $\al,\al^*$ is a couple of complex-conjugated 
 (Fock-like)  variables 
[see later on in Eq. (\ref{etatime}) for a relationship between these 
variables, the space-time position $x^\mu$ and  the covariant momenta 
$p^\mu$, corresponding 
to  the  more usual  ``configuration-space'' image].

The group law $U''=U'U$ in $\eta,\al,\al^*$ coordinates, 
\begin{eqnarray}
\eta'' &=&\frac{z_1''}{|z_1''|}=\frac{\eta'\eta+\eta'\eta^*\al'\al^*}{\sqrt{(1+
{\eta^*}^2\al'\al^*)(1+\eta^2\al{\al^*}')}}\,,  \nn \\
\al''&=&\frac{z_2''}{z_1''}=\frac{\al\eta^2+\al'}{\eta^2+\al'\al^*}\,,
\label{law1} \\
{\al^*}''&=&\frac{{z_2''}^*}{{z_1''}^{*}}=
\frac{\al^*\eta^{-2}+{\al^*}'}{\eta^{-2}+{\al^*}'\al}\;,  \nn
\end{eqnarray}
\ni can be centrally-extended by 
$U(1)$ through a cocycle (in fact, coboundary) 
generated by the function $\delta(g)=-2iNy_0$ on 
the basic group $G$, in the following form:

\be
\zeta''=\zeta'\zeta
e^{i\xi(g',g)}=\zeta'\zeta\left(\eta''{\eta'}^{-1}\eta^{-1}\right)^{-2N}\;;
\;\;\zeta, \zeta' \in U(1)\,,\label{extension1}        
\ee
\ni where the parameter $N$ appears to be quantized for globallity 
conditions [compactness of the variable $\zeta$ and $\eta$; $2N$ is the 
``winding number'' of applications from $U(1)\subset SU(1,1)\rightarrow
 U(1)$], the possible values 
for $N$ being $N=\frac{n}{2},\,n\in Z$, and characterizes each irreducible 
representation of $\TG$. The integrallity condition for $2N$ disappears on the 
universal covering CAdS of AdS space-time.

The explicit expression of the left- and right-invariant vector fields of 
the extended group $\TG$ is:
\bea
\TXL_\zeta&=&\TXR_\zeta=\zeta\f{\p}{\p\zeta}\label{campos} \\
\TXL_\eta &=&\eta\f{\p}{\p\eta}-2\al\f{\p}{\p \al}+2\al^*
\f{\p}{\p \al^*} \nn \\
\TXL_\al &=&-\f{1}{2}\eta \al^* \f{\p}{\p\eta}+\f{\p}{\p \al}-
{\al^*}^2\f{\p}{\p \al^*}+N\alc\zeta\f{\p}{\p\zeta}\nn \\
\TXL_{\al^*} &=&\f{1}{2}\eta \al \f{\p}{\p\eta}-\al^2 \f{\p}{\p \al}+
\f{\p}{\p \al^*}-N\al\zeta\f{\p}{\p\zeta}\nn \\
\TXR_\eta &=&\eta \f{\p}{\p\eta} \nn \\
\TXR_\al &=& \f{1}{2}\eta^{-1} \al^* \f{\p}{\p\eta}+\eta^{-2}(1-\al\al^*)
\f{\p}{\p \al}-N\eta^{-2}\alc\zeta\f{\p}{\p\zeta}\nn \\
\TXR_{\alc} &=&- \f{1}{2}\eta^3 \al \f{\p}{\p\eta}+\eta^{2}(1-\al\al^*)
\f{\p}{\p\alc}+N\eta^{2}\al\zeta\f{\p}{\p\zeta}\, .\nn
\eea
\ni Both sets close the Lie algebra:
\bea
\l[\TXL_\eta,\TXL_\al\r]&=& 2\TXL_\al \nn \\
\l[\TXL_\eta,\TXL_{\alc}\r] &=& -2\TXL_{\alc} \nn \\
\l[\TXL_\al,\TXL_{\alc}\r] &=& \TXL_\eta -2N\TXL_\zeta\nn \\
\l[\TXL_\zeta,\hbox{all}\r]&=&0\;, \label{conmutadores1}
\eea
\ni and the corresponding right counterparts change a sign in the 
structure constants. 
The Casimir operator for this Lie algebra is  (except for a central term 
ambiguity):
\be
\hat{C}=(\TXR_\eta-2N\TXR_\zeta)^2+2\TXR_\al\TXR_{\alc}+
2\TXR_{\alc}\TXR_\al=-(\TXR_{y_0}-2N\TXR_{y_3})^2+(\TXR_{y_1})^2+
(\TXR_{y_2})^2 \label{casimir}\;,
\ee
\ni where we have denoted $\al\equiv y_1+iy_2$ and $\zeta\equiv e^{iy_3}$ 
for future use. 

The Quantization 1-form and the Characteristic Module 
have 
the following form:
\bea
\Theta&=&\f{iN}{1-\al\alc}\left(4\al\alc\eta^{-1}d\eta+\alc 
d\al-\al
d\alc\right)-i\zeta^{-1}d\zeta \nn \\
{\cal G}_\Theta &=& <\TXL_\eta>\; .
\eea

We realize that a full polarization (first-order and complete) exist 
for this holomorphic picture and it is made up of
\be
{\cal P}=<\TXL_\eta,\TXL_\al>\; . \label{pola1}
\ee
\ni The solution of the $T$-function condition (\ref{tcondition}), 
together with the polarization conditions (\ref{defpol}), lead to the 
following wave functions:
\bea
{\psi^{(N)}}(\eta,\al,\alc,\zeta)&=& W_N(\al,\alc,\zeta)\phi(s) \nn \\
W_N(\al,\alc,\zeta)&=&\zeta (1-\al\alc)^{N}\, ,       \label{wavepol}
\eea
\ni where $W_N$ plays the role of a vacuum and $\phi$ 
is an arbitrary power series
\be
\phi(s)=\sum_{n=0}^{\infty}a_n s^n,
\ee
\ni in the variable
\be 
s\equiv\eta^{-2}\alc=\f{{z_2^*}}{{z_1}}. 
\ee
The group $\TG$ acts on this one-dimensional space by left translation 
($\tg''=\tg'*\tg$) as follows:  
\be
s\rightarrow s'=\f{s+\alc{}'}{{\eta'}^2(1+s \al')}\, .
\ee

\ni The left-invariant integration volume has the form  
\bea
\mu(\al,\alc,\eta,\zeta)&=&\f{-(2\pi)^{-3}}{(1-\al\alc)^2}
\eta^{-1}d\eta\wedge 
d\hbox{Re}(\al)\wedge d\hbox{Im}(\al)\wedge
\zeta^{-1}d\zeta\,. \label{volume}
\eea
\ni Let us denote  
$\check{\psi}^{(N)}_n(\eta,\al,\alc,\zeta)=
W_N(\al,\alc,\zeta)s^n$ a basic 
wave function. The scalar product of two of them is:
\be
\langle \check{\psi}^{(N)}_n | \check{\psi}^{(N)}_m\rangle = 
\f{n!(2N-2)!}{(2N+n-1)!}\delta_{nm}\equiv C^{(N)}_n\delta_{nm},
\ee
\ni which is well-defined (finite) for values $N > \f{1}{2}$. This 
condition can be relaxed to $N>0$ by going to the universal covering group 
of $G$. The set
\be 
B({\cal H}_N(\TG))=\left\{ \psi^{(N)}_n\equiv\f{1}{\sqrt{C^{(N)}_n}}
\check{\psi}^{(N)}_n \right\}\label{basicwave}
\ee  
is then orthonormal and complete, i.e, an orthonormal base of 
${\cal H}_N(\TG)$.

The action of the right-invariant vector fields 
(operators in the theory) on
polarized wave functions in (\ref{wavepol}) has the explicit form:
\bea
\TXR_\eta{\psi^{(N)}} &=& W_N\cdot(-2s\f{\p}{\p s})\phi(s)\,, \nn \\
\TXR_\al{\psi^{(N)}} &=& W_N\cdot(-s^2\f{\p}{\p s}-2Ns)\phi(s)\,,\nn\\
\TXR_{\alc}{\psi^{(N)}} &=& W_N\cdot(\f{\p}{\p s})\phi(s)\,,\nn\\
\TXR_\zeta{\psi^{(N)}}&=&{\psi^{(N)}}, \label{repreos}
\eea
\ni providing an action on the $\phi$-space once the common factor
$W_N$ has been factored out.  

The action of the Casimir operator 
(\ref{casimir}) on polarized wave functions is:
\be
\hat{C}\psi^{(N)}=4N(N-1)\psi^{(N)}\label{casimiract}\;.
\ee

The  finite (left) action (\ref{leftaction}) of an arbitrary element 
$\tg'=\tg'(\eta',\al',\alc{}',\zeta')\in\TG$ on an arbitrary wave function 
\be
\psi^{(N)}(\tg)=\sum^\infty_{n=0}a_n\psi^{(N)}_n(\tg)\;,
\ee
\ni can be given through the matrix elements 
$\rho^{(N)}_{mn}(\tg')\equiv\langle \psi^{(N)}_m|\rho(\tg')|
\psi^{(N)}_n\rangle$ of $\rho$ in the base $B({\cal H}_N(\TG))$. They have 
the following expression:
\bea
\rho^{(N)}_{mn}(\tg)&=& \sqrt{\f{C^{(N)}_m}{C^{(N)}_n}}\zeta^{-1}
\sum^n_{l={\rm Max}(0,n-m)}
\left(\begin{array}{c}n\\ l\end{array}\right)
\left(\begin{array}{c}2N+m+l-1\\ m-n+l\end{array}\right)\times \nn \\
  & &(-1)^l\eta^{2m}{\alc}^l\al^{m-n+l}(1-\al\alc)^N\;.
\eea

Let us show how to second-quantize this first-quantized theory.

\section{Quantum Field Theory on a symmetric curved space}

\subsection{``Second Quantization'' on a group}

In this subsection we shall develop a general approach to the 
quantization of  linear, complex quantum fields defined on a group 
manifold $\TG$ [more precisely, on the quotient $\TG/(T\cup P)$]. 
This formalism can be seen as a ``second quantization" of a ``first-quantized" 
theory defined by a group $\TG$ and a Hilbert space ${\cal H}(\TG)$ of 
polarized wave functions. 

The construction of the quantizing group 
$\tilde{G}^{(2)}$ for this complex quantum field is as follows. Given 
a  Hilbert space ${\cal H}(\TG)$ and its dual ${\cal H}^*(\TG)$, we 
define the direct sum 
\bea
{\cal F}(\TG)&\equiv& {\cal H}(\TG)\oplus {\cal H}^*(\TG)\nn \\
&=&\left\{|f\rangle =|A\rangle +|B^*\rangle ;\,\,
|A\rangle \in {\cal H}(\TG),\,|B^*\rangle \in {\cal H}^*(\TG)\right\}\;,
\eea
\ni where we have denoted  $|B^*\rangle$ according to 
  $\langle \tg_\cp^*|B^*\rangle \equiv 
\langle B|\tg_\cp\rangle =B_\cp^*(\tg)$. The group $\TG$ acts on this 
vectorial space as follows:
\be
\rho(\tg')|f\rangle =\rho(\tg')|A\rangle +\rho(\tg')|
B^*\rangle\;, 
\ee
\ni where 
\be
\langle \tg_\cp^*|\rho(\tg')|B^*\rangle \equiv 
\langle B|\rho^{\dag}(\tg')|\tg_\cp\rangle =B_\cp^*(\tg'{}^{-1}*\tg). 
\ee

We can also define the dual space 
\bea
{\cal F}^*(\TG)&\equiv& {\cal H}^*(\TG)\oplus {\cal H}^{**}(\TG)\nn \\
&=&\left\{\langle f|=\langle A|+\langle B^*|\,;\;\;
\langle A|\in {\cal H}^*(\TG),\,\langle B^*|\in {\cal H}^{**}(\TG)\sim 
{\cal H}(\TG)\right\}\;,
\eea
\ni where $\TG$ acts according to the adjoint action
\be
\langle f|\rho^{\dag}(\tg')=\langle A|\rho^{\dag}(\tg')+
\langle B^*|\rho^{\dag}(\tg')
\ee
\ni and now 
\be
\langle B^*|\rho^{\dag}(\tg')|\tg_\cp^*\rangle \equiv
\langle \tg_\cp|\rho(\tg')|B\rangle .
\ee

Using the closure relation (\ref{closure}), the product of two arbitrary 
elements of ${\cal F}(\TG)$ is
\be
\langle f'|f\rangle =\langle A'|A\rangle +
\langle A'|B^*\rangle +
\langle B'{}^*|A\rangle +
\langle B'{}^*|B^*\rangle=\langle A'|A\rangle+\langle B'{}^*|B^*\rangle \,,
\ee
\ni since the integrals 
\be
\int_{\TG}{\mu(\tg)A_\cp'{}^*(\tg)B_\cp^*(\tg)}=0=
\int_{\TG}{\mu(\tg)B_\cp'(\tg)A_\cp(\tg)}
\ee
\ni are zero because of the integration on 
the central parameter $\zeta\in U(1)$. Thus, the subspaces 
${\cal H}(\TG)$ and ${\cal H}^*(\TG)$ are orthogonal with respect to this 
scalar product in ${\cal F}(\TG)$. A basis for ${\cal F}(\TG)$ is provided by 
the set  ${\cal B}({\cal F}(\TG))=\left\{|n\rangle + 
|m^*\rangle \right\}_{n,m\in I}\,$.

The space ${\cal M}(\TG)\equiv{\cal F}(\TG)\otimes {\cal F}^*(\TG)$ 
can be endowed with a symplectic structure
\be
S(f',f)\equiv\f{-i}{2}(\langle f'|f\rangle -
\langle f|f'\rangle )\,,
\ee
\ni thus defining  ${\cal M}(\TG)$ as 
a phase space. This symplectic structure in ${\cal M}(\TG)$ is paralleled by 
the symplectic structure, $\Omega$, in the solution manifold ${\cal S}$ of 
the classical equations (\ref{K}) in other approaches to QFT in globally 
hyperbolic curved space (see, for example, \cite{Wald2}), or by the 
canonical product $\Omega$ in Ref. \cite{config2}.

This phase space can be naturally 
embedded into a quantizing group 
\be
\tilde{G}^{(2)}\equiv\left\{\tg^{(2)}=(g^{(2)};\vs)=\left(\tg,|f\rangle ,
\langle f|;\vs\right)\right\}\;,
\ee

\ni which is a (true) central extension by $U(1)$, with parameter $\vs$, of 
the semidirect product $G^{(2)}\equiv\TG\otimes_{\rho}{\cal M}(\TG)$ of 
the basic group $\TG$ and the phase space ${\cal M}(\TG)$. The group 
law of $\tilde{G}^{(2)}$ is formally:
\bea
\tg''&=&\tg'*\tg \nn \\
|f''\rangle &=&|f'\rangle +\rho(\tg')|f\rangle \nn\\
\langle f''|&=&\langle f'|+\langle f|\rho^{\dag}(\tg')\nn\\
\vs''&=&\vs'\vs e^{i\xi^{(2)}(g^{(2)}{}',\,g^{(2)})}\;,\label{group2}
\eea
\ni where $\xi^{(2)}(g^{(2)}{}',g^{(2)})$ is a cocycle defined as
\be
\xi^{(2)}(g^{(2)}{}',g^{(2)})\equiv \kappa S(f',\rho(\tg')f)
\label{coci2}
\ee
\ni and $\kappa$ is intended to kill any possible dimension of $S$. The 
group law (\ref{group2}) generalizes in a natural way the semi-direct 
action of the time evolution uniparametric group by the 
Heisenberg-Weyl group modeled 
on the solution manifold of a linear field, which has the structure of 
an infinte-dimensional symplectic vector space. Although the representation 
$\rho(\tg)$ of the space-time symmetry group $\TG$ on the fields $f$ could be 
more general (even more, $\tilde{G}^{(2)}$ might admit other  
structure than the semi-direct product), 
we shall restrict ourselves in this paper to the left-finite action 
(\ref{leftaction}) coming from a first quantization on the group $\TG$, thus 
leading to a ``second quantization on a group $\TG$'' paralleling the 
standard concept in QFT of ``second quantization''. 
The unitarity of $\rho$ ensures the hermiticity of the space-time symmetry 
operators (in particular, the Hamiltonian), something which is guaranteed 
when $\rho$ comes from a first quantization on $\TG$. 

A system of coordinates for $\tilde{G}^{(2)}$ corresponds to a choice of 
representation associated with a given polarization ${\cal P}$
\be
\begin{array}{ll} f_\cp^{(+)}(\tg)\equiv\langle \tg_\cp|f\rangle \,, & 
f_\cp^{(-)}(\tg)\equiv\langle \tg_\cp^*|f\rangle \,,\\
f_\cp^{*(+)}(\tg)\equiv\langle f|\tg_\cp^*\rangle \,, &
f_\cp^{*(-)}(\tg)\equiv\langle f|\tg_\cp\rangle \,.\end{array}
\label{confpar}
\ee
\ni This splitting of $f$ is the group generalization of the more standard 
decomposition of a field in positive and negative frequency parts. 
If we make use of the closure relation 
$1=\int_{\TG}{ \mu(\tg)\{|\tg_\cp\rangle\langle \tg_\cp|+|\tg_\cp^*\rangle  
\langle \tg_\cp^*|\}}$ for ${\cal F}(\TG)$, the explicit 
expression of the cocycle (\ref{coci2}) in this coordinate system 
(for simplicity, we discard the semidirect action of $\TG$),  
\bea 
\xi^{(2)}(g^{(2)}{}',g^{(2)})&=&\f{-i\kappa}{2}\int\!\!\int_{\TG}{
\mu(\tg')\mu(\tg) \left\{ 
f_\cp'^{*(-)}(\tg')\Delta_\cp^{(+)}(\tg',\tg)f_\cp^{(+)}(\tg)\right.} \nn\\ 
&-&  f_\cp^{*(-)}(\tg')\Delta_\cp^{(+)}(\tg',\tg)f_\cp'^{(+)}(\tg) 
  + f_\cp'^{*(+)}(\tg')\Delta_\cp^{(-)}(\tg',\tg)f_\cp^{(-)}(\tg)
\label{coci22} \\ 
&-&  \left. f_\cp^{*(+)}(\tg')\Delta_\cp^{(-)}(\tg',\tg)f_\cp'^{(-)}(\tg) 
\right\}\,,\nn
\eea
\ni where 
\bea
\Delta_\cp^{(+)}(\tg',\tg)&\equiv&\langle\tg_\cp'|\tg_\cp\rangle=\sum_{n\in I}{
\psi_{\cp,n}(\tg')\psi_{\cp,n}^*(\tg)}\,,\nn\\
\Delta_\cp^{(-)}(\tg',\tg)&\equiv&\langle\tg_\cp'^*|\tg_\cp^*\rangle=
\Delta_\cp^{(+)}(\tg,\tg')\,,\label{propm}
\eea
\ni shows that the vector fields associated with  the co-ordinates in 
(\ref{confpar}) are canonically conjugated
\bea
\l[\TXL_{f^{*(-)}_\cp(\tg')},\TXL_{f_\cp^{(+)}(\tg)}\r]&=&\kappa
\Delta_\cp^{(+)}(\tg',\tg)\TXL_\vs\,,\nn\\
\l[\TXL_{f^{*(+)}_\cp(\tg')},\TXL_{f_\cp^{(-)}(\tg)}\r]&=&\kappa
\Delta_\cp^{(-)}(\tg',\tg)\TXL_\vs\,.
\eea
\ni Here, the functions $\Delta_\cp^{(\pm)}(\tg',\tg)$ play the role of  
{\it propagators} 
(central matrices of the cocycle). The propagators in two different 
parametrizations of $\tilde{G}^{(2)}$, corresponding to two different 
polarization subalgebras ${\cal P}_1$ and ${\cal P}_2$ of ${\tilde{\cal G}}^L$ 
(or $U{\tilde{\cal G}}^L$),  
are related through polarization-changing operators (\ref{operchan}) as 
follows:
\bea
\Delta_{\cp_2}^{(\pm)}(\th',\th)&=&\int\!\!\int_{\TG}{\mu(\tg')\mu(\tg)
\Delta_{\cp_2\cp_1}^{(\pm)}(\th',\tg')\Delta_{\cp_1}^{(\pm)}(\tg',\tg)
\Delta_{\cp_1\cp_2}^{(\pm)}(\tg,\th)}\,,\nn\\
\Delta_{\cp_i\cp_j}^{(+)}(\th,\tg)&\equiv&\Delta_{\cp_i\cp_j}(\th,\tg)\,,
\;\;\;\Delta_{\cp_i\cp_j}^{(-)}(\th,\tg)\equiv\Delta_{\cp_i\cp_j}(\tg,\th)
\,.\label{polchanop}
\eea
\ni A particular case of the last expression is the transformation 
which relate the propagator $\langle p'|p\rangle$ in momentum-space 
to the propagator $\langle x'|x\rangle$ in configuration-space.

At this point, it is important to stress that a globally hyperbolic structure 
($\Re\times\Sigma$) in the space $Q=G/P$ is not a prerequisite 
for having a well-defined quantum field theory constructed from the group  
$\tilde{G}^{(2)}$. The necessity of a spatial hypersurface 
$\Sigma\subset Q$ ---on which 
Cauchy data are freely specified in the standard 
approach--- to quantize a field  
on $Q$ is now inessential because  we can perform the quantization program  
in an alternative representation space (momentum $p$-space,  
Bargmann-Fock $\alpha^*$-space, etc) for which the possible undesirable 
global properties of $Q$ might be absent. Even for the quantization in 
configuration space, it should be stressed that the existence of 
$\Sigma$ (i.e., the necessity of a well-posed 
{\it classical\,}  initial value formulation) is not the problem that really 
matters in GAQ but, rather, the existence of an appropriate polarization 
intended to reduce unitarily the group representation, as already commented. 
Note that we are computing inner products by integrating on the whole group 
$\TG$, not just on $\Sigma$. There may be some cases where we could 
factorize out the integration on the time parameter in 
$\int_{\TG}{\mu(\tg)}$, together with 
the extra (momentum) variables (by following 
a regularization process in the case in which the co-ordinates are  
non-compact), keeping an
 integration $\int_\Sigma d\sigma(x)$ on $\Sigma\subset Q$ only; 
however, this is not the general case, and an integration on the time 
parameter (even on the whole group $\TG$) could 
be  necessary to keep hermiticity in the symmetry generators [remember  
the comments after (\ref{unitpoin}) and see later on in Sec. 4].

As regard the general problem of the ``back-reaction'' effects of 
the quantum field on the curved space-time \cite{Wald3}, 
it might also be worthmentioning  
the {\it Hadamard} condition on states of 
the first-quantized theory as a necessary condition for a state to have a 
finite expected value for the 
stress-energy tensor \cite{Wald3,Wald4}. For Hadamard states, 
two-point functions (the analog of our $\Delta^{(+)}_\cp 
+\Delta^{(-)}_\cp$) satisfy a certain asymptotic condition \cite{Kay2}. 
Nevertheless, the proper discussion of this general 
problem in a group framework 
requires the stress-energy tensor to be incorporated as an infinite set 
of extra generators (related to a gauge-like algebra)  
in the group ${\tilde G}^{(2)}$ of the second quantized theory. 
This enlargement of ${\tilde G}^{(2)}$ is beyond the scope of this article; 
however, a treatment of this kind has been performed in connection with 
2d-Gravity \cite{hidden}.

In applying the GAQ formalism to $\tilde{G}^{(2)}$, it is appropriate to 
use a ``Fourier-like" parametrization asociated with the basis 
${\cal B}({\cal F}(\TG))=\left\{|n\rangle + 
|m^*\rangle \right\}_{n,m\in I}\,$ made of Hamiltonian eigenstates, 
alternative to the field-like parametrization above 
[see (\ref{confpar})]\footnote{Of course, a ``manifestly covariant'' 
parametrization $f(x)=\langle x|f\rangle$ in configuration-space 
relative to boost-like eigenstates 
$|\tg_\cp=x\rangle$ is also possible \cite{config2}, 
although the characterization 
of the Polarization subalgebra (see bellow) is much 
more involved as it makes use of the manifestly covariant propagator 
$\Delta(x,x')=\langle x|x'\rangle$ (see e.g. Eq. (\ref{propaholo})), 
versus the more manageable 
$\Delta(n,n')=\langle n|n'\rangle=\delta_{n,n'}$.}. If we denote by
\be
\begin{array}{ll} a_n\equiv\langle n|f\rangle \,, & 
b_m\equiv\langle m^*|f\rangle \,,\\ a^*_n\equiv\langle f|n\rangle \,, & 
b^*_m\equiv\langle f|m^*\rangle \,,\end{array}\label{fupa}
\ee
\ni the Fourier coefficients of the ``particle" and the ``antiparticle", 
 a polarization subalgebra ${\cal P}^{(2)}$ 
for $\tilde{G}^{(2)}$ can always be given by  
\be
{\cal P}^{(2)}=<\,\TXL_{a_n},\TXL_{b_m}; \,\,
{{\cal G}}_{\Theta^{(2)}}\simeq\tilde{{\cal G}}^L\,>\,,\;\;\forall n,m\in I,
\label{pol2nd}
\ee
that is, the corresponding left-invariant vector fields 
$\TXL_{a_n},\TXL_{b_n}$ and 
the whole Lie algebra $\tilde{{\cal G}}^L$ of $\TG$, which is the 
characteristic subalgebra 
${{\cal G}}_{\Theta^{(2)}}$ of 
the second-quantized theory (see next subsection). The operators of the 
theory are the right-invariant vector fields of $\tilde{G}^{(2)}$; 
in particular, the basic operators are: the annihilation operators of 
particles and anti-particles,  $\hat{a}_n\equiv\TXR_{a^*_n} ,\,
\hat{b}_m\equiv\TXR_{b^*_m}$, and the corresponding creation operators 
 $\hat{a}^\dag_n\equiv-\frac{1}{\kappa}\TXR_{a_n} ,\,\hat{b}^\dag_m\equiv 
-\frac{1}{\kappa}\TXR_{b_m}$. The operators corresponding to the subgroup 
$\TG$ [the second-quantized version $\TXRS_{\tg^j}$ of the first-quantized 
operators $\TXR_{\tg^j}$ in (\ref{txlr1})] are written in terms of the basic 
ones (it is worthmentioning that they appear, in a natural way, {\it normally 
ordered}), since they are in the characteristic subalgebra 
${{\cal G}}_{\Theta^{(2)}}$ of the second-quantized theory.

The group $\TG$ plays a key role in characterizing vacuum states in the 
curved space $Q$, 
in the same way as the Poincar\'e 
group plays a central role in relativistic quantum theories defined on 
Minkowski space. 
In general, standard QFT in curved space suffers from the lack of a preferred 
definition of particles. The infinite-dimensional character of the symplectic 
solution manifold of a field system is responsible for the existence of an 
infinite number of unitarily inequivalent irreducible representations
of the Heisenberg-Weyl (H-W) relations and there 
is no criterion to select a preferred vacuum of the corresponding quantum 
field  (see, for example, \cite{Birrell,Wald2,Segal,Berezin}). 
This situation is not present in the finite-dimensional case, 
according to the Stone-von Neuman theorem \cite{Sharon,von}. 
In our language, the origin of this fact is related to the infinite, 
arbitrary, non-equivalent directions that the wave functions 
$\Psi(a,a^*,b,b^*)$ can be polarized for the 
infinite-dimensional H-W subgroup ${\tilde G}^{(2)}/\TG$ 
of ${\tilde G}^{(2)}$ itself.  In practice, this means that arbitrary 
choices of annihilation operators, through the canonical 
transformation (Bogolyubov transformation)
\be
\hat{a}'_l=\sum_{n\in I}\alpha_{ln}\hat{a}_n+\beta_{ln}\hat{a}^\dag_n\,,\;\;\;
\hat{a}'^\dag_l=\sum_{n\in I}\bar{\beta}_{ln}\hat{a}_n+
\bar{\alpha}_{ln}\hat{a}^\dag_n\,
\label{Bogolyubov}
\ee
(and similarly for the antiparticle) 
lead to non-identical (non-unitarily equivalent) vacua charactericed by 
$\hat{a}_n|0\rangle=0$ and 
$\hat{a}'_l|0'\rangle=0\,\,,\forall n,l\in I$, when $\beta_{ln}$ is not 
a Hilbert-Schmidt operator (see \cite{Berezin}).  
The situation is rather different, however, when we can embed the curved 
space $Q$ into a given group $\TG$. In fact, the existence 
of a characteristic module
---generated by ${{\cal G}}_{\Theta^{(2)}}\sim {\tilde{\cal G}}^L$---  
in the polarization subalgebra ${\cal P}^{(2)}$ strongly restricts the 
possible transformations (\ref{Bogolyubov}) to automorphysms of 
${\cal P}^{(2)}$, that is, the transformation (\ref{Bogolyubov}) has now to 
preserve the maximality and horizontality properties characterizing 
a full polarization subalgebra like (\ref{pol2nd}). In other words, the 
transformation (\ref{Bogolyubov}) has to be one-to-one 
(in order for ${\cal P}^{(2)}$ to be maximal), and to fulfill: 
\be
\l[\TXLS_{\tg^j},\TXL_{a'_l}\r]=\rho'^{j}_{lk}\TXL_{a'_k},\,\,\,\forall 
\tg\in\TG,\,\,\forall l\in I\,,
\ee
in order to close a horizontal subalgebra, where the transformed 
structure constants $\rho'^{j}_{lk}$ must be related to the original 
$\rho^{j}_{mn}\equiv \left.\partial\rho_{mn}(\tg)/\partial\tg^j
\right|_{\tg=e}$ by 
\be
\rho'^{j}_{lk}=\alpha_{lm}\rho^{j}_{mn}\alpha^{-1}_{nk}=\beta_{lm}
\rho^{j*}_{mn}\beta^{-1}_{nk}\,.
\ee
The last equation imposes strong restrictions to the coeficients 
$\alpha,\beta$, which also have to fulfill the typical relations 
for an infinite-dimensional symplectic transformation (see e.g. 
\cite{Birrell,Wald2,Segal,Berezin}), 
leaving a limited number of possible polarization 
subalgebras ${\cal P}^{(2)}$. For each ${\cal P}^{(2)}$, the corresponding 
vacuum state will be characterized as being annihilated 
by the right version of the 
polarization subalgebra dual to ${\cal P}^{(2)}$. For example, 
in the case of (\ref{pol2nd}), the vacumm will be 
invariant under the action of 
$\TG\subset {\tilde G}^{(2)}$ and annihilated by 
the  right-invariant vector fields $\TXR_{a^*_n},\TXR_{b^*_n}$ (the standard 
annihilation operators $\hat{a}_n,\,\hat{b}_n$, respectively). 
In more physical terms, the vacuum states are those 
which ``look the same" to any  freely falling observer, anywhere in the 
curved space $Q=G/P$. 

Moreover, it is also always possible to choose 
particular states which behave as vacua with respect 
to a given subgroup $\TG_K\subset \TG$; that is,  those states which are  
invariant under $\TG_K$ only. For example, $\TG_K$ could be 
the uniparametric subgroup of time evolution (see e.g. \cite{Bruce} for 
a discussion of vacuum states in de Sitter space). Extremely interesting 
are the physical phenomena related to the choice of Weyl (Poincar\'e + 
dilatation) invariant pseudo-vacua (zero-mode coherent states) in a 
conformally invariant QFT. Vacuum radiation in relativistic 
acceletated frames and Fulling-Unruh effect \cite{Fulling,Unruh} 
are discussed in this framework in Ref. \cite{conforme}.

>From a formal point of view, the choice of particular pseudo-vacua 
would correspond to a breakdown of the symmetry and could be 
understood as a constrained version of the original theory. 
Indeed, let us  comment on the 
influence of the constraints 
in the first-quantized theory at the second quantization level. 
Associated with
a constrained wave function satisfying (\ref{const}), there is a 
corresponding constrained quantum field subjected to the condition:
\be
\hbox{ad}_{\TXRS_{\tilde{\tau}}}\left(\TXR_{|f\rangle }\right)\equiv
\l[\TXRS_{\tilde{\tau}},\TXR_{|f\rangle }\r]=
dD^{(\epsilon)}_T(\tilde{\tau})\TXR_{|f\rangle }\,,
\ee
\ni where  $\TXRS_{\tilde{\tau}}$ stands for the ``second-quantized 
version" of $\TXR_{\tilde{\tau}}$. It is straightforward to generalize 
the last condition to higher-order constraints: 
\bea
\TXR_1\TXR_2...\TXR_j|\psi\rangle &=&\epsilon |\psi\rangle 
\rightarrow\nn\\
\hbox{ad}_{\TXRS_1}\left(\hbox{ad}_{\TXRS_2}\left(...\hbox{ad}_{\TXRS_j}\left(
\TXR_{|f\rangle }\right)...\right)\right)&=&\epsilon\TXR_{|f\rangle }\,.
\label{goodfield}
\eea
\ni  The selection of a given Hilbert subspace 
${\cal H}^{(\epsilon)}(\TG)\subset {\cal H}(\TG)$ made of wave 
functions obeying a higher-order constraint $K\psi=\epsilon\psi$, where 
$K=\TXR_1\TXR_2...\TXR_j$ is some Casimir operator of $\TG_K\subset \TG$,  
manifests itself, at second quantization level, as a new 
(broken) QFT. The vacuum for the new observables of this broken theory 
(the good operators in (\ref{goodfield})) does not have to coincide with 
the vacuum of the original theory, and the action of the rest of the operators 
(the bad operators) could make this new vacuum  radiate (see Ref. 
\cite{conforme}).

In general, constraints lead to gauge symmetries 
in the constrained theory and, also, 
the property for a subgroup $N\subset \TG$ 
of being gauge is inheritable at the second-quantization level.

To conclude this subsection, it is important 
to note that the representation of $\TG$ on ${\cal M}(\TG)$ is reducible, but 
it is irreducible under $\TG$ together with the {\it charge conjugation} 
operation $a_n\leftrightarrow b_n$, which could be implemented on 
$\tilde{G}^{(2)}$. For simplicity, we have preferred to discard this 
transformation; 
however, a treatment including it, would be relevant as a revision of 
the CPT symmetry in quantum field theory. The Noether invariant associated 
with $\TXRS_\zeta$ is nothing other than the {\it total electric charge} 
(the total number of particles 
in the case of a real field 
$b_n\equiv a_n$) and its central character, inside the 
``dynamical" group $\TG$ of the first-quantized theory, now ensures its 
conservation under the action of the subgroup $\TG\subset\TG^{(2)}$. 
To account for non-Abelian charges (iso-spin, color, etc),
 a non-Abelian structure group 
$T\subset\TG$ is required.

\subsection{Example of the AdS space}

Let us apply  the GAQ formalism to the centrally extended 
group $\TG^{(2)}$ given through the group law in 
(\ref{group2}) for the case of $G=SO(2,1)$ and the ``holomorphic 
polarization"  used in Subsec. \ref{holosec}. 
For simplicity, we shall consider 
the case of a real field and we shall use the   ``Fourier" parameterization 
(\ref{fupa}) in terms of the coefficients $a_n$ rather than the ``field"  
parameterization (\ref{confpar}) 
in terms of $f_\cp^{(\pm)}(\tg)$. The explicit group law is:
\bea
\tg'' &=&\tg'*\tg \label{halr}\\
{a''_m}&=&{a'_m}+\sum^\infty_{n=0}\A_{mn}(\tg')a_n \nn\\
{a^*_m}''&=&{a^*_m}'+\sum^\infty_{n=0}{\A_{mn}}^*(\tg')a^*_n \nn\\
\vs''&=&\vs'\vs\exp{\f{\kappa}{2}\sum^\infty_{m=0}
\sum^\infty_{n=0}({a^*_m}'
\A_{mn}(\tg')a_n-{a'_m}{\A_{mn}}^*(\tg')a^*_n)}\,.\nn
\eea

Let us denote $\p_m\equiv\f{\p}{\p a_m},\;
\p^*_m\equiv\f{\p}{\p a^*_m}$ and use the parameterization 
$\tg(\eta,\al,\alc,\zeta)=\tg(y_\nu), \nu=0,1,2,3$ after (\ref{ycero}) 
and (\ref{casimir}), in order to 
deal with hermitian operators $\TXRS_{y_\nu}$. 
With this notation, the left- and right-invariant vector fields are:
\bea
\TXL_\vs&=&\TXR_\vs=\vs\f{\p}{\p \vs}\nn \\
\TXL_{a_n}&=&\sum^\infty_{m=0}\A_{mn}(y){\p_m}+\f{\kappa}{2}
\sum^\infty_{m=0}\A_{mn}(y)a^*_m\TXL_\vs\nn \\
\TXL_{a^*_n}&=&\sum^\infty_{m=0}{\A_{mn}}^*(y){\p^*_m}-\f{\kappa}{2}
\sum^\infty_{m=0}{\A_{mn}}^*(y)a_m\TXL_\vs\nn \\
\TXLS_{y_\nu}&=&\TXL_{y_\nu}\nn\\
&{}&\nn\\
\TXR_{a_n}&=&{\p_n}-\f{\kappa}{2}a^*_n\TXL_\vs\nn \\
\TXR_{a^*_n}&=&{\p^*_n}+\f{\kappa}{2}a_n\TXL_\vs\nn \\
\TXRS_{y_0}&=&\TXR_{y_0}+2i\sum^\infty_{m=0}m(a_m{\p_m}-
a^*_m{\p^*_m})\nn\\
\TXRS_{y_1}&=&\TXR_{y_1}+\sum^\infty_{m=0}\sqrt{(m+1)(2N+m)}
(a_m{\p_{m+1}}-a_{m+1}{\p_m}+
a^*_m{\p^*_{m+1}}-a^*_{m+1}{\p^*_m})\nn\\
\TXRS_{y_2}&=&\TXR_{y_2}+i\sum^\infty_{m=0}\sqrt{(m+1)(2N+m)}
(a_m{\p_{m+1}}+a_{m+1}{\p_m}-
a^*_m{\p^*_{m+1}}-a^*_{m+1}{\p^*_m})\nn\\
\TXRS_{y_3}&=&\TXR_{y_3}-i\sum^\infty_{m=0}(a_m{\p_m}-
a^*_m{\p^*_m})\,.
\eea
The non-trivial commutators among 
those vector fields are:
\bea
\l[\TXLS_{y_0},\TXLS_{y_1}\r]&=&-2\TXLS_{y_2}\nn\\
\l[\TXLS_{y_0},\TXLS_{y_2}\r]&=&2\TXLS_{y_1}\nn\\
\l[\TXLS_{y_1},\TXLS_{y_2}\r]&=&-2\TXLS_{y_0}+4N\TXLS_{y_3}\nn\\
\l[\TXLS_{y_0},\TXL_{a_n}\r]&=&2in\TXL_{a_n}\nn\\
\l[\TXLS_{y_1},\TXL_{a_n}\r]&=&-\sqrt{n(2N+n-1)}\TXL_{a_{n-1}}
+\sqrt{(n+1)(2N+n)}\TXL_{a_{n+1}}\nn\\
\l[\TXLS_{y_2},\TXL_{a_n}\r]&=&i\sqrt{n(2N+n-1)}\TXL_{a_{n-1}}
+i\sqrt{(n+1)(2N+n)}\TXL_{a_{n+1}}\nn\\
\l[\TXLS_{y_3},\TXL_{a_n}\r]&=&-i\TXL_{a_n}\nn\\
\l[\TXLS_{y_0},\TXL_{a^*_n}\r]&=&-2in\TXL_{a^*_n}\nn\\
\l[\TXLS_{y_1},\TXL_{a^*_n}\r]&=&-\sqrt{n(2N+n-1)}\TXL_{a^*_{n-1}}
+\sqrt{(n+1)(2N+n)}\TXL_{a^*_{n+1}}\nn\\
\l[\TXLS_{y_2},\TXL_{a^*_n}\r]&=&-i\sqrt{n(2N+n-1)}\TXL_{a^*_{n-1}}
-i\sqrt{(n+1)(2N+n)}\TXL_{a^*_{n+1}}\nn\\
\l[\TXLS_{y_3},\TXL_{a^*_n}\r]&=&i\TXL_{a^*_n}\nn\\
\l[\TXL_{a_n},\TXL_{a^*_m}\r]&=&-\kappa\delta_{nm}\TXL_{\vs}.
\eea

The quantization 1-form  and the characteristic module are:
\bea
\Theta^{(2)}&=&\f{i\kappa}{2}\sum^\infty_{n=0}
(a_nda^*_n-a^*_nda_n)-i\vs^{-1}d\vs \nn\\
{{\cal G}}_{\Theta^{(2)}}&=&<\TXLS_{y_\nu}>,\,\, 
\nu=0,1,2,3.\label{2form}
\eea
\ni The Noether invariants of the second-quantized theory are:
\bea
F_{a_n}&=&i_{\TXR_{a_n}}\Theta^{(2)}=-i\kappa a^*_n\nn\\
F_{a^*_n}&=&i_{\TXR_{a_n}}\Theta^{(2)}=i\kappa a_n\nn\\
F^{(2)}_{y_0}&=&i_{\TXRS_{y_0}}\Theta^{(2)}=
2\kappa\sum^\infty_{n=0}na^*_na_n\nn\\
F^{(2)}_{y_1}&=&i_{\TXRS_{y_1}}\Theta^{(2)}=i\kappa\sum^\infty_{n=0}
\sqrt{(n+1)(2N+n)}(a^*_na_{n+1}-a^*_{n+1}a_n)\nn\\
F^{(2)}_{y_2}&=&i_{\TXRS_{y_2}}\Theta^{(2)}=\kappa\sum^\infty_{n=0}
\sqrt{(n+1)(2N+n)}(a^*_na_{n+1}+a^*_{n+1}a_n)\nn\\
F^{(2)}_{y_3}&=&i_{\TXRS_{y_3}}\Theta^{(2)}=-\kappa\sum^\infty_{n=0}
a^*_na_n\,.\label{2noether}
\eea

A full polarization subalgebra can be:
\be
{\cal P}^{(2)}=<\TXLS_{y_\nu},\TXL_{a_n}>, \;\;\forall n\geq 0,\;\;
\nu=0,1,2,3\,,\label{polal}
\ee
\ni and the polarized $U(1)$-functions have the form:
\be
\Psi[a_n,a^*_n,y_\nu,\vs]=\vs\exp\left\{-\f{\kappa}{2}\sum^\infty_{n=0}a^*_na_n
\right\}\Phi[a^*]\equiv \Omega\Phi[a^*]\,,\label{2wavepol}
\ee
\ni where $\Omega$ is the vacuum of the second quantized 
theory and $\Phi$ is an arbitrary power series in its arguments. As 
already commented, $\Omega$ looks the same to any freely falling observer 
anywhere in the AdS space. It is annihilated by the  right version 
of the dual of (\ref{polal}) as can be seen from the general  action of the 
right-invariant vector fields 
(operators in the second-quantized theory) on
polarized wave functions in (\ref{2wavepol}). This action has 
the explicit form: 
\bea
\TXR_{a_n}\Psi&=&\Omega\cdot(-\kappa a^*_n)\Phi\equiv 
\Omega\cdot(-\kappa {\hat{a}}^{\dag}_n)\Phi\nn\\
\TXR_{a^*_n}\Psi&=&\Omega\cdot(\p^*_n)\Phi\equiv 
\Omega\cdot({\hat{a}}_n)\Phi\nn\\
\TXRS_{y_0}\Psi&=&\Omega\cdot\left(-2i\sum^\infty_{n=0}n\ac_n\a_n\right)
\Phi\equiv-i\Omega
{\hat{F}}^{(2)}_{y_0}\Phi\nn\\
\TXRS_{y_1}\Psi&=&\Omega\cdot\left(\sum^\infty_{n=0}\sqrt{(n+1)(2N+n)}
(\ac_n\a_{n+1}-\ac_{n+1}\a_n)\right)\Phi\equiv-i\Omega{\hat{F}}^{(2)}_{y_1}
\Phi\nn\\
\TXRS_{y_2}\Psi&=&\Omega\cdot\left(-i\sum^\infty_{n=0}\sqrt{(n+1)(2N+n)}
(\ac_n\a_{n+1}+
\ac_{n+1}\a_n)\right)\Phi\equiv-i\Omega{\hat{F}}^{(2)}_{y_2}\Phi\nn\\
\TXRS_{y_3}\Psi&=&\Omega\cdot\left(i\sum^\infty_{n=0}{\ac_n\a_n}\right)
\Phi\equiv-i\Omega{\hat{F}}^{(2)}_{y_3}\Phi\;,
\eea
\ni where $\a_n$ and $\ac_n$ are interpreted as annihilation and creation 
operators, ${\hat{F}}^{(2)}_{y_0}$ is interpreted as the total energy operator 
(Hamiltonian), ${\hat{F}}^{(2)}_{y_3}$ represents the total number 
of particles (the total electric charge in the complex case),  
 and the remainder corresponds to other 
conserved quantities of the theory. As already said, all those 
quantities appear, in a natural way, {\it normally ordered}. This is one 
of the advantages of this method of quantization: normal order does not have 
to be imposed by hand but, rather, it is implicitly inside the 
formalism.

Let us go back to the expression (\ref{2form}) of the Quantization 
1-form. Note that its simple appearance  is due to the fact that 
it is written in terms of the initial 
condition variables (\ref{2noether}) masking the dynamical 
content of it. Let us perform a change of variables induced by a general action
of the group $\TG$ (\ref{halr})
\be
a_n=\sum^\infty_{m=0}\A_{nm}(y)c_m\,,
\ee
\ni and express $\Theta^{(2)}$ in terms of the ``evolving" variables 
$c_m$. The final form of it is as follows:
\bea
\Theta^{(2)}&=&\f{i\kappa}{2}\sum^\infty_{n=0}
(c_ndc^*_n-c^*_ndc_n+T^\nu_n(y)dy_\nu)-i\vs^{-1}d\vs \nn\\
T^\nu_n(y)&\equiv&\sum^\infty_{m=0}\sum^\infty_{l=0}\left(\A_{nm}(y)
\f{\p{\A_{nl}}^*(y)}{\p y_\nu}-{\A_{nl}}^*(y)\f{\p \A_{nm}(y)}{\p y_\nu}
\right)c^*_l c_m.
\eea
\ni The quantities $T^\nu_n(y)$ play the role of momenta (for instance, 
$T^0_n(y)$ is the $n$-mode energy). 

For completeness, we shall give the explicit expression of the propagator 
(\ref{propm})  in the present holomorphic polarization case. After some 
calculations,  it proves to be:
\be
\Delta^{(+)}(\tg',\tg)=\sum^{\infty}_{n=0}{\psi^{(N)}_n(\tg')
\psi^{(N)*}_n(\tg)}
=(2N-1)\frac{\zeta'\zeta^*(1-\al'{\alc}')^N(1-\al\alc)^N}{(1-{\alc}'\al
\eta'{}^{-2}\eta^2)^{2N}}\,.\label{propaholo}
\ee
In the configuration-space image, the corresponding propagator 
can be calculated 
by making use of the expression (\ref{polchanop}) and the 
polarization-changing operator given in \cite{A-B-G-N} (``relativistic 
Bargmann transform'').  
Let us discuss a bit more on the 
connection between the ``holomorphic picture" used through the paper 
and the more usual ``configuration-space picture" used in standard approaches 
to QFT, within the GAQ framework.

\section{Configuration-space image}

The Quantum Mechanics on AdS can be also built in the position 
representation,  which is the usual image when one makes use of the 
intrinsic differentiable structure of this space-time. In order to make 
contact with this standard approach, 
let us consider the 3D hyperboloid defining the de Sitter and 
AdS universes (see e.g. \cite{Weinberg,Hawking})
 
 \begin{equation}
 \eta_{\mu\nu}x^{\mu}x^{\nu}+\frac{\lambda^{2}}{k}=\frac{1}{k}\;,
 \label{hyperbo}
 \end{equation}
 
 \noindent where $\eta_{\mu\nu}=(+,-), x^{\mu}=(x^0,x^1)$, $\lambda$ is the 
extra co-ordinate and $k$ stands for the curvature of the space. For $k>0$, 
Eq. (\ref{hyperbo}) defines the AdS 
space-time ($SO(1,2)$ homogeneous space), for $k<0$, it defines the 
de Sitter space-time ($SO(2,1)$ homogeneous space), and for $k=0$, we get the 
Minkowski space-time.
 
 The line element in the flat 3-D space defined by (\ref{hyperbo}) is

\begin{equation}
 c^{2}(d\tau)^{2}=\eta_{\mu\nu}dx^{\mu}dx^{\nu}+\frac{1}{k}(d\lambda)^{2} ,
 \label{line}
 \end{equation}
 
 \noindent where $\tau$ stands for the proper time. We solve for 
$\lambda$ in
 (\ref{hyperbo}) and differentiate,
 
 \begin{eqnarray}
 \lambda&=&\sqrt{1-k\eta_{\mu\nu}x^{\mu}x^{\nu}} \nn \\
 d\lambda&=&-\frac{k\eta_{\mu\nu}x^{\mu}dx^{\nu}}{\lambda} , \label{dy}
 \end{eqnarray}
 
 \noindent and, introducing (\ref{dy}) into (\ref{line}),
 
 \begin{equation}
 c^{2}(d\tau)^{2}=\eta_{\mu\nu}dx^{\mu}dx^{\nu}+
\frac{k}{\lambda^{2}}x^{\rho}
 \eta_{\rho\mu}x^{\sigma}\eta_{\sigma\nu}dx^{\mu}dx^{\nu}\;,
 \label{lineindu}
 \end{equation}
 
 \noindent we obtain the metric induced by (\ref{line}) in the 
tangent space to the
 hyperboloid (\ref{hyperbo}); this is the de Sitter metric:
 
 \begin{equation}
 g_{\mu\nu}=\eta_{\mu\nu}+k\frac{\eta_{\mu\lambda}x^{\lambda}
\eta_{\nu\kappa}x^{\kappa}}
 {1-k\eta_{\rho\sigma}x^{\rho}x^{\sigma}} ,
 \label{indumetric}
 \end{equation}
 
 \noindent with inverse
 
 \begin{equation}
 g^{\mu\nu}=\eta^{\mu\nu}-kx^{\mu}x^{\nu} .
 \end{equation}
 
 Let us define the contravariant momenta for a particle of mass m,
 $p^{\mu}=m\frac{dx^{\mu}}{dt}$. Then, Eq. (\ref{lineindu}) may be
 regarded as a constraint between the momenta $p^{\mu}$, i.e. the 
 de Sitter mass shell
 
 \begin{equation}
 (p^{0})^{2}-(p)^{2}+\frac{k}{\lambda^{2}}(p^{0}x^0-px^1)^{2}=m^{2}c^{2} .
 \label{deSittermass}
 \end{equation}
 
 \noindent Solving $p^{0}$ in terms of $p$ and $x^{\mu}$, we get
 
 \begin{equation}
 p^{0}=\frac{kpx^1x^{0}+\lambda\sqrt{m^{2}c^{2}+p^{2}+
m^{2}c^{2}k{x^{1}}^2}}{1+k{x^{1}}^2} .
 \label{p0deSitter}
 \end{equation}
 
 From (\ref{indumetric}) it follows the connection
 
 \begin{equation}
 \Gamma^{\mu}_{\nu\lambda}=kx^{\mu}g_{\nu\lambda} \, ,
 \end{equation}
 
 \noindent so that the geodesics are given by the solutions of the
 equation
 \begin{equation}
 \frac{d^{2}x^{\mu}}{d\tau^{2}}=-kx^{\mu} ,\label{geodesic}
 \end{equation}
 \ni which could also be interpreted as the equation of motion of 
a (general) relativistic oscillator \cite{A-B-G-N}. 

To give the action of the (anti-)de Sitter 
group $G$ over space-time, we shall adopt the prescription that 
AdS-parameters become those of the Poincar\'e group action under the 
limit $k\rightarrow 0$. By denoting $a^0,a^1,p$ the AdS space-time 
translation parameters and boost parameters, respectively, as well 
as $\Lambda$ the group version of the extra co-ordinate $\lambda$, the above 
mentioned  group action proves to be:
\begin{equation}
 \left(\begin{array}{c}
      {x'}^{0}\\{x'}^{1}\\ \lambda'\end{array}\right)=
        \left(\begin{array}{ccc}
      \frac{p^{0}}{mc}&\frac{p}{mc}&a^{0}\\
      \frac{p}{mc}&\frac{p^{0}}{mc}&a^{1}\\
-k\frac{p^{0}a^{0}-pa^{1}}{\Lambda mc}&kK&\Lambda\end{array}\right)
    \left(\begin{array}{c}x^{0}\\x^{1}\\ \lambda\end{array}\right)\,,
\label{matrixaction}
\end{equation}
\ni  where we have defined
 \begin{equation}
 K\equiv\frac{p^{0}a^{1}-pa^{0}}{mc};\;\;
\Lambda\equiv\sqrt{1-k({a^{0}}^{2}-{a^{1}}^{2})}\,,
 \end{equation}
\ni and $p^0$ is defined in terms of $a^0,a^1,p$ as in (\ref{p0deSitter}). 
>From the action (\ref{matrixaction}) it is straightforward to derive
the group law by letting it act twice ($g''=g'*g$). The explicit form of 
this group law is:
\begin{eqnarray}
 a^{0}{}''&=&\frac{{p'}^{0}}{mc}a^{0}+\frac{P'}{mc}a^{1}+
        \Lambda a^{0}{}' \nn \\
 a^{1}{}''&=&\frac{p'}{mc}a^{0}+\frac{P^{0}{}'}{mc}a^{1}+
       \Lambda a^{1}{}'  \label{law}  \\
 p''&=&\frac{p'p^{0}}{mc}+\frac{P^{0}{}'p}{mc}-
       \frac{k}{\Lambda}(p^{0}a^{0}-pa^{1})a^{1}{}' ,\nn
 \end{eqnarray}
\ni where we have defined
\begin{equation}
P^{0}\equiv\frac{p^{0}+mcka^{1}K}{\Lambda};
\;P\equiv\frac{p+mcka^{0}K}{\Lambda}\,.
\end{equation}

It is a very convenient practice to chose a compact time parameter in 
order to work out the quantum dynamics of a system with bound states. If 
we denote 
\begin{equation}
t\equiv\frac{1}{\omega}\arcsin\frac{\omega a^0}{c\beta};\;\;
\beta\equiv\sqrt{1+kx^2};\;\;\omega\equiv\sqrt{k}c;\;x\equiv a^1
\end{equation}
\ni and then consider $g=g(t,x,p)$ a parameterization of $G$, a 
pseudo-extension by $U(1)$ (i.e. a cocycle generated by a function 
$\delta(g)$ on the basic group, usually linear in time) can be 
chosen to be:
\begin{eqnarray}
 \zeta''&=&\zeta'\zeta e^{\frac{i}{\hbar}(\delta(g'')-
\delta(g')-\delta(g))}\,,\nn \\
\delta(g)&\equiv&-mc^2\left(t+\frac{1}{mc^2}f(x,p)\right)\,, \nn\\
 f(x,p)&\equiv&-\frac{2mc^{2}}{\omega}\arctan\left[\frac{mc^{2}}
{\omega px}(\beta-1)
 (\frac{P_{0}}{mc}-\beta)\right]\,, \nn\\ 
P_{0}&\equiv&\sqrt{mc^{2}+p^{2}+m^{2}\omega^{2}x^{2}}\,.\label{extencur} 
 \end{eqnarray}
The new coordinates $x, p, t$ are related to the old coordinates 
$\al, \alc, \eta$ as follows:
\begin{eqnarray}
 \sqrt{\frac{m\omega}{2\hbar}}\,x+\frac{i}{\sqrt{2m\hbar\omega}}\,p
&\equiv &\sqrt{2N}\f{\al}{1-\al\alc}\,,  
\label{etatime} \\
\exp{-i\frac{\w}{2mc^2}\delta}&\equiv &\eta=\exp{iy_0}\,, \nn
\end{eqnarray}

\ni where the extension parameter $N$ proves to be 
$N=\frac{mc^2}{\hbar\omega}$, and suffers from the same requirements as before
[see the comment after (\ref{extension1})].  The function $f(x,p)$ generalizes 
the simple expression 
$f(x,p)\sim xp$, for the flat and non-relativistic case, to the curved case.

The commutation relations among left-invariant vector fields 
associated  with the co-ordinates $t,x,p,\zeta$ of $\TG$ are:
\begin{eqnarray}
 \left[\tilde{X}^{L}_{t}, \; \tilde{X}^{L}_{x}\right]&=&
-mkc^{2}\tilde{X}^{L}_{p} \nn \\
 \left[\tilde{X}^{L}_{t}, \; \tilde{X}^{L}_{p}\right]&=&
\frac{1}{m}\tilde{X}^{L}_{x} \nn \\
 \left[\tilde{X}^{L}_{x}, \; \tilde{X}^{L}_{p}\right]&=&
\frac{1}{mc^{2}}\tilde{X}^{L}_{t}-i\frac{1}{\hbar}\TXL_\zeta \nn \\
 \left[\tilde{X}^{L}_{\zeta}, \; \hbox{all}\right]&=&0\, ,
\label{conmurelas}
 \end{eqnarray}
\ni and the quantization 1-form and its characteristic module are:  
\begin{eqnarray}
\Theta &=& \frac{p\left(P_0^2+m^2\w^2 x^2\right)}
{\beta^2P_0\left(P_0+mc\right)}dx -
         \frac{m^2c^2x}{P_0\left(P_0+mc\right)}dp - 
c\left(P_0-mc\right)dt +
          \hbar\frac{d\zeta}{i\zeta}  \nn\\
{\cal G}_\Theta&=&<\TXL_t>\label{tetarelconf} \, .
\end{eqnarray}
\ni The Noether invariants are immediately calculated 
 as $i_{\tilde{X}^{R}}\Theta$:
\begin{eqnarray}
i_{\tilde{X}^R_t}\Theta &=& -c(P_0-mc)\equiv E \nn \\
i_{\tilde{X}^R_x}\Theta &=& \frac{p}{\beta}\cos\omega t + 
            \frac{P_0}{mc\beta} m\omega x\sin\omega t \equiv P \\
i_{\tilde{X}^R_p}\Theta &=& -\frac{P_0}{mc\beta} x\cos\omega t+
        \frac{p}{m\omega\beta}\sin\omega t \equiv -K \nn \,,
\end{eqnarray}
\ni which fulfill the classical mass-shell restriction
 \begin{equation}
 E^{2}-m^{2}c^{2}\omega^{2}K^{2}-c^{2}P^{2}=m^{2}c^{4} .
 \label{CMSC}
 \end{equation}

Let us now consider  the set ${\cal B}(\TG)$ of complex valued 
$U(1)$-functions on $\TG$. As we have already mentioned, the 
representation of $\TG$ on ${\cal B}(\TG)$ is reducible. The reduction 
can be achieved by means of polarization conditions (\ref{defpol}). We 
seek  an explicit representation of $\TG$ on wave functions 
defined on the AdS space-time, this implying that the generator 
$\TXL_p$ (generator corresponding to the subgroup $P$) has to be included 
in the polarization. The commutation relations (\ref{conmurelas}) force 
us to ``deform" the characteristic module $\TXL_t$ to a ``higher order" 
$\tilde{X}^{HO}_{t}$ operator, which we can choose to be basically the 
left version of the Casimir operator plus an arbitrary 
central term; more explicitly:
\begin{equation}
{\cal P}^{HO}= <\tilde{X}^{HO}_{t}\equiv(\tilde{X}^{L}_{t})^{2}-c^{2}
(\tilde{X}^{L}_{x})^{2}-\frac{2imc^{2}}{\hbar}
    \tilde{X}^{L}_{t}-\frac{mc^{2}\omega}{\hbar}\TXL_\zeta,
\,\tilde{X}^{L}_{p}> \, ,
 \label{GHOP}
 \end{equation}
 
 \noindent  The $U(1)$-function condition (\ref{tcondition}) together 
with the polarization conditions lead to wave functions which fulfill:

 \begin{eqnarray}
 \TXL_\zeta\psi=\psi&\rightarrow&\psi=\zeta\phi(x,p,t) \nn \\
 \tilde{X}^{L}_{p}\psi=0&\rightarrow&\psi=
\zeta e^{\frac{i}{\hbar}(f+mc^2)}\varphi(x,t) \nn  \\
 \tilde{X}^{HO}_{t}\psi=0&\rightarrow&\left(\square_x + 
\frac{m^2c^2}{\hbar^2} + \nu R\right)\varphi = 0\, , \label{KGEqcurv}
\end{eqnarray}
\ni where $R\equiv -2k$ can be viewed as the scalar curvature (see 
\cite{Birrell} for instance), $\nu\equiv\f{N}{2}$ and 
\be
\square_x\equiv\frac{1}{c^{2}\beta^{2}}\frac{\partial^{2}}
{\partial t^{2}}-
      \frac{2\omega^{2}x}{c^{2}}\frac{\partial}{\partial x}-
      \beta^{2}\frac{\partial^{2}}{\partial x^{2}} \, ,\label{Dalam}
\ee
\ni is the D' Alambertian operator on AdS space-time. Clearly, 
for $k\rightarrow 0$, 
equation (\ref{KGEqcurv}) goes to the usual Klein-Gordon 
equation for a free particle moving on Minkowski space-time.

The general equation (\ref{KGEqcurv}) can be solved by power-series 
expansion, leading to  functions
 \begin{equation}
 \varphi_n\equiv e^{-iE^{(N)}_n\omega t}
\beta^{-E^{(N)}_n}
\HNln\, ,\label{wavege}
 \end{equation}
\ni where $E^{(N)}_n=N+n$,  and 
$\HNln(\chi)$ are polynomials in the variable 
$\chi\equiv\sqrt{\f{m\omega}{\hbar}}x$, which can be written 
via the Rodrigues formula ($\lambda=1$ in Ref. \cite{A-B-G-N}) as
\be
\HNln(\chi)=(-1)^n\left(1+\frac{\chi^2}{N}\right)^{N+n}\frac{d^n}{d\chi^n}
\left[\left(1+\frac{\chi^2}{N}\right)^{-N}\right]\,.
\ee
There is a polarization-changing operator $\langle x|\alpha\rangle$ 
(called ``relativistic 
Bargmann transform" and explicitly calculated in \cite{A-B-G-N}) 
relating wave functions (\ref{wavege}) in configuration space representation 
to those given in (\ref{basicwave}) as corresponding to a holomorphic 
(Fock-like) representation. The polynomials $\HNln$ are related 
to the Gegenbauer \cite{Gegenbauer}  and  Jacobi \cite{Ismail} polynomials.   
The wave functions (\ref{wavege}) reproduce, for the 1+1D case,  
those found in  \cite{Ishamdisrup} (see also the 
2nd paper in Ref. \cite{Fronsdal}) 
for the massive case, provided that we perform 
the change of variables $x\rightarrow \rho\equiv 
\hbox{arctan}(\frac{\omega}{c}x), \,\,\,t\rightarrow \tau\equiv 
\frac{\pi}{2}-\omega t$. In fact, except for normalization constants and 
time dependence, our even wave functions can be written as 
\bea
\varphi_{2n}&\sim& \beta^{-N}\beta^{-2n}H^{(N)}_{2n}(\chi)
\nn\\ 
&\sim& \beta^{-N}P^{(n-\frac{1}{2},-\frac{1}{2})}_n\left(
\frac{\chi^2-N}{\chi^2+N}\right)\sim (\cos\rho)^N
P^{(-\frac{1}{2},N-\frac{1}{2})}_n(\cos 2\rho)\,,
\eea
which correspond to the 1+1D version of the wave functions in Ref. 
\cite{Ishamdisrup}. The restriction to even wave functions is simply 
because we are comparing  1+1D wave functions with the radial 
part of the 3+1D ones. Note also that, in 1+1D, the range of $M=N$ 
would be $M=1,2,3,\dots$ instead of $M=3,4,5\dots$ [$N$ is half-integer 
when working in the two-covering of AdS group $SO(1,2)$].  

The invariant integration volume in $\TG$ now adopts the following form:
\be
\mu(\tg)=-i\f{dp\wedge dx\wedge dt\wedge d\zeta}{\zeta P_0}\,,\label{esc1}
\ee
\ni respect to which the group $\TG$ is unitarily represented in the 
present configuration-space. Thus, quantization in a highly symmetric 
non-globally hyperbolic space-time $Q$, such as AdS, can proceed on  
the basis of symmetry 
by translating the problem of existence of a well-defined, 
deterministic classical evolution in the standard (canonical) quantization 
scheme, to a problem of unitarily reducing the representations of  
$\TG\supset Q$ in the GAQ framework. In general, unitarity implies 
self-adjointness of the generators of the group and, in particular, 
of the time generator, leading to an inherently conserved Noether 
invariant (the energy) in GAQ. Both properties are not always ensured 
in the canonical quantization on non-globally 
hyperbolic spaces, unless additional boundary conditions are imposed 
\cite{Ishamdisrup,Peter}.

As mentioned above, the reduction 
$\int_{\TG}{\mu(\tg)}\rightarrow\int_{\Sigma}{d\sigma(x)}$ to a minimal 
canonical representation of $\TG$ on $\Sigma\subset Q$ 
could lead, 
in general,  to a loss of hermiticity of some part of the operators in 
${\tilde{\cal G}}^R$. Indeed, 
we must stress the different structure of the time evolution in Minkowski  
space in Subsec. \ref{frp} (or, in general, globally hyperbolic space-times)  
as compared 
with the AdS case. The time parameter cannot be 
factorized out in a natural way. The appearance of the partial weights 
$\beta^{-n}$ in
the wave functions in configuration space 
is traced back to
the presence of a time derivative term in the quantum operators. 
Another consequence of the structure of time evolution (manifest 
covariance of our configuration space representation) is the need for the time 
integration in the scalar product defined through the left-invariant 
integration volume in (\ref{esc1}). In fact, a naive 
factorization 
of the time dependence in the wave functions, operators and scalar 
product leads to a non-unitary representation; the functions 
in (\ref{wavege})  are no longer 
orthogonal nor the operators $\TXR_x,\TXR_p$ are hermitian. The 
$x$-representation can nevertheless be 
``unitarized'' by changing the integration measure, $dx 
\rightarrow 
\frac{dx}{\beta^2}$, and by redefining the operators  $\TXR_x$ and $\TXR_p$ 
accordingly. The redefinition process really
parallels the {\it multipliers method} used in the literature 
\cite{Hermann,Bargmann2,Mackey} to
construct unitary representations of a group G when a natural 
invariant
volume is absent. The measure $\frac{dx}{\beta^2}$ is left 
invariant under
the $U(1)$ subgroup of $SO(1,2)$, i.e. the time evolution, so that the 
energy operator is not affected by the multipliers.  Furthermore, 
the space-time parts of our wave
 functions $\varphi(x,t)$ satisfy a Klein-Gordon-like equation 
(\ref{KGEqcurv})
 with an operator $\square_x$ (\ref{Dalam}) associated with the metric
 $ds^2=c^2 \beta^2 dt^2 -\beta^{-2}dx^2$ and, according to the 
standard
 techniques in Quantum Mechanics, one would define the 
time-invariant scalar product
 $\int\frac{dx}{\beta^2}\varphi^*\varphi'$ in order to have a 
conserved probability
 density current. The new representation obtained by this way is a 
minimal representation and it really constitutes a 
well-defined theory of orthogonal functions (polynomials with 
partial weights; see e.g. \cite{A-B-G-N,Gegenbauer}). 

We should remark that, even though the Bargmann-Fock-like representation 
(\ref{wavepol}, \ref{repreos}) can be
directly restricted to the $\al,\al^*$-dependence without losing unitarity,
a time-independent polarization-changing operator is not directly defined.

\section*{Acknowledegment}

M. Calixto thanks the University of Granada for a Post-doctoral grant and the 
Department of Physics of Swansea for its hospitality.




\begin{thebibliography}{99}

\bibitem{Fulling} S. A. Fulling, Phys. Rev. {\bf D7}, 2850, (1973).
\bibitem{Birrell}       N.D. Birrell and P.C.W. Davies, {\it 
         Quantum fields in curved space},
                       Cambridge University Press, Cambridge (1982).
\bibitem{Wald2} R.M. Wald, {\it Quantum Fields in Curved Space and Black
Hole Thermodynamics}, University of Chicago Press, (1995).
\bibitem{Hawking2} S.W. Hawking, Commun. Math. Phys. {\bf 43}, 199 (1975).
\bibitem{Unruh} W. G. Unruh, Phys. Rev. {\bf D14}, 870 (1976).
\bibitem{conforme} V. Aldaya, M. Calixto and J.M. Cerver\'o, 
Commun. Math. Phys. {\bf 200}, 325 (1999).
\bibitem{Kirillovb} A. A. Kirillov, {\it Elements of the Theory of 
                  Representations}, Springer-Verlag (1976).
\bibitem{GQ1}           J.M. Souriau, {\it Structure des systemes 
                         dynamiques},
                         Dunod, Paris (1970).
\bibitem{GQ2}           B. Kostant, {\it Quantization and Unitary
                         Representations},
                         in Lecture Notes in Math. {\bf 170}, 
                         Springer-Verlag, Berlin (1970).
\bibitem{GQ3}           J. Sniatycki, {\it Geometric Quantization 
                         and Quantum Mechanics},
                         Springer-Verlag, New York (1970).
\bibitem{GQ4}         N. Woodhouse, {\it Geometric Quantization},
                         Clarendon, Oxford (1980).
\bibitem{Kirillov} A. A. Kirillov, {\it Geometric Quantization}, in Dynamical 
Systems IV (Symplectic Geometry and its Applications), V.I. Arnol'd and 
S.P. Novikov (Eds.), Springer Verlag (1989).
\bibitem{GAQ}  V. Aldaya and J. de Azc\'arraga, 
                 J. Math. Phys. {\bf 23}, 1297 (1982)
\bibitem{AGAQ}  V. Aldaya, J. Navarro-Salas and A. Ram\'\i rez,  Commun. 
                   Math. Phys. {\bf 121}, 541 (1989). 
\bibitem{pertur} J. Guerrero and V. Aldaya, Mod. Phys. Lett. {\bf A14},
           1689 (1999).
\bibitem{oscipert} V. Aldaya and J. Guerrero, {\it Perturbative-group 
  quantization} (in preparation).
\bibitem{Fulling2} S. A. Fulling, {\it Aspects of Quantum Field 
Theory in Curved Space-Time}, Cambridge University Press, Cambridge, (1989).
\bibitem{Waldjmp} Robert M. Wald, J. Math. Phys. {\bf 21}, 2802 (1980).
\bibitem{chorri} V. Aldaya, J. Navarro-Salas, J. Bisquert and R. Loll,
                 J. Math. Phys. {\bf 33}, 3087-3097 (1992).
\bibitem{Marmo} V. Aldaya, J. Guerrero and G. Marmo, Int. J. Mod. Phys. 
         {\bf A12}, 3 (1997).
\bibitem{Hawkingtmu} S. A. Hawking, Phys. Rev. {\bf D46}, 603, (1992).
\bibitem{Kay1} B. S. Kay, M. J. Radzikowski and R. M. Wald, 
         Commun. Math. Phys. {\bf 183}, 533, (1997).
\bibitem{Peter} Peter Breitenlohner and Daniel Z. Freedman, Ann. Phys. 
{\bf 144}, 249 (1982).
\bibitem{FracHall} V. Aldaya, M. Calixto and J. Guerrero, 
        Commun. Math. Phys. {\bf 178}, 399 (1996).
\bibitem{virazorro} V. Aldaya and J. Navarro-Salas, 
           Commun. Math. Phys. {\bf 139}, 433 (1991).
\bibitem{Extensiones} V. Bargmann, Ann. Math. {\bf 59}, 1 (1954).
\bibitem{Saletan} E. J. Saletan,  J. Math. Phys. {\bf 2}, 1 (1961).
\bibitem{Pseudoco} V. Aldaya and J.A. de Azc\'arraga, Int. J. Teor. Phys. 
        {\bf 24}, 141 (1985).
\bibitem{Marmo2} V. Aldaya, J. Guerrero, G. Marmo, {\it Quantization 
of a Lie Group: Higher Order Polarizations},  in "Symmetries in Science X", 
Ed. Bruno Gruber and Michael Ramek, Plenum Press New York (1998).
\bibitem{Abraham} R. Abraham and J. E. Marsden, 
{\it Foundations of Mechanics}, 
W. A. Benjamin, Inc., Reading, Massachusetts, (1967).
\bibitem{config2}  M. Navarro, V. Aldaya and M. Calixto, 
                J. Math. Phys. {\bf 38}, 1454 (1997); 
                J. Math. Phys. {\bf 37}, 206,(1996).
\bibitem{Hermann}    R. Hermann, {\it Lie groups for physicists}, 
                      W.A. 
                     Benjamin, INC., New York (1966). 
\bibitem{Jackiw} R. Jackiw and C. Rebbi, Phys. Rev. Lett. 
{\bf 37} 172 (1976)\\ 
C. Callan, R. Dashen and D. Gross, Phys. Lett. {\bf B63} 334 (1976).
\bibitem{Position} V. Aldaya, J. Bisquert, J. Guerrero and J. Navarro-Salas, 
                     J. Phys. {\bf A 26}, 5375  (1993) 
\bibitem{Mir-Kasimov3} R.M. Mir-Kasimov, J. Phys. {\bf A 24}, 4283 (1991)
\bibitem{spoincare} V. Aldaya and J. Guerrero, J. Phys. {\bf A28}, L137 (1995).
\bibitem{Hawking}       S.W. Hawking and G.F. Ellis, {\it The Large
                 Scale Structure of Space-Time}, 
                Cambridge University Press, Cambridge (1973).
\bibitem{Weinberg}      S. Weinberg, {\it Gravitation and 
                 Cosmology}, Wiley, New York (1972).
\bibitem{Wald3} Robert M. Wald, Commun. Math. Phys. {\bf 54}, 1 (1977).
\bibitem{Wald4} Robert M. Wald, Phys. Rev. {\bf D17}, 1477 (1978).
\bibitem{Kay2} B. S. Kay and R. M. Wald, Phys. Rep. {\bf 207}, 49, (1991).
\bibitem{hidden} V. Aldaya, J. Navarro-Salas and M. Navarro, J. Phys. 
{\bf A26}, 5391 (1993);  
Contemporary Mathematics {\bf 132}, 1 (1992).  
\bibitem{Sharon} M. Stone, Proc. Nat. Ac. (1929, 1930).
\bibitem{von} J. von Neumann, Math. Ann. Bd. 102 (1929).
\bibitem{Segal} I.E. Segal, {\it Mathematical Problems of Relativistic 
Physics}, Am. Math. Soc., Providence, R.I. (1963).
\bibitem{Berezin} F.A. Berezin, {\it The Method of Second Quantization}, 
Academic, New York (1966).
\bibitem{Bruce} Bruce Allen, Phys. Rev.{\bf D32}, 3136 (1985).
\bibitem{A-B-G-N} V. Aldaya, J. Bisquert, J. Guerrero and J. Navarro-Salas,
 Reports on Math. Phys.{\bf 37}, 387-417 (1995). 
\bibitem{Gegenbauer}  B. Nagel, J. Math. Phys. {\bf 35}, 1549 (1994).
\bibitem{Ismail} Mourad E. H. Ismail, J. Phys. {\bf A29}, 3199 (1996).
\bibitem{Ishamdisrup} S. J. Avis, C. J. Isham and D. Storey, Phys. Rev. 
{\bf  D18}, 3565 (1978).
\bibitem{Fronsdal} C. Fronsdal, Rev. Mod. Phys., {\bf 37}, 221 (1965).\\  
 C. Fronsdal, Phys. Rev. {\bf D10}, 589 (1974).\\  C. Fronsdal, 
Phys. Rev. {\bf D12}, 3819 (1975). \\ 
C. Fronsdal and R. B. Haugen, Phys. Rev. {\bf D12} 
3810 (1975). 
\bibitem{Bargmann2}  V. Bargmann, Ann. Math. {\bf 48}, 568 (1947). 

\bibitem{Mackey}     G.W. Mackey, {\it Unitary Group Representations
                     in Physics, Probability, and Number Theory}, 
                     Mathematics
                     Lecture Notes Series, Benjamin/Cummings 1978; 
                     S. Lang, 
                $SL_2(R)$, Addison-Wesley Publishing Company 1975.
\end{thebibliography}
\end{document}